\documentclass[twocolumn]{aastex631}

\shorttitle{Variable SN 2023ixf progenitor}
\shortauthors{Soraisam et al.}

\begin{document}

\title{The SN 2023ixf Progenitor in M101: I. Infrared Variability}

\correspondingauthor{Monika Soraisam}
\email{monika.soraisam@noirlab.edu}

\author[0000-0001-6360-992X]{Monika~D.~Soraisam}
\affiliation{Gemini Observatory/NSF’s NOIRLab, 670 N.~A’ohoku Place, Hilo, HI 96720, USA}
\author[0000-0003-4610-1117]{Tam\'as~Szalai}
\affiliation{Department of Experimental Physics, Institute of Physics, University of Szeged, D\'om t\'er 9, 6720 Szeged, Hungary}
\affiliation{ELKH-SZTE Stellar Astrophysics Research Group, Szegedi {\'u}t, Kt.~766, 6500 Baja, Hungary}
\author[0000-0001-9038-9950]{Schuyler~D.~Van~Dyk}
\affiliation{Caltech/IPAC, Mailcode 100-22, Pasadena, CA 91125, USA}
\author[0000-0003-0123-0062]{Jennifer~E.~Andrews}
\affiliation{Gemini Observatory/NSF’s NOIRLab, 670 N.~A’ohoku Place, Hilo, HI 96720, USA}
\author[0000-0002-2996-305X]{Sundar~Srinivasan}
\affiliation{Instituto de Radioastronom{\'i}a y Astrof{\'i}sica, UNAM, Antigua Carretera a P\'atzcuaro 8701, Ex-Hda. San Jos\'e de la Huerta,  Morelia 58089, Mich., Mexico}
\author[0000-0002-6154-7558]{Sang-Hyun~Chun}
\affiliation{Korea Astronomy and Space Science Institute, 776 Daedeokdae-ro, Yuseong-gu, Daejeon 34055, Republic of Korea}
\author[0000-0001-6685-0479]{Thomas~Matheson}
\affiliation{NSF's NOIRLab, 950 N.~Cherry Ave, Tucson, AZ 85719, USA}
\author[0000-0002-1161-3756]
{Peter~Scicluna}
\affiliation{European Southern Observatory, Alonso de Cordova 3107, Santiago RM, Chile}
\author[0009-0008-2354-0049]
{Diego A. Vasquez-Torres}
\affiliation{Instituto de Radioastronom{\'i}a y Astrof{\'i}sica, UNAM, Antigua Carretera a P\'atzcuaro 8701, Ex-Hda. San Jos\'e de la Huerta,  Morelia 58089, Mich., Mexico}

\begin{abstract}
Observational evidence points to a red supergiant (RSG) progenitor for SN~2023ixf. The progenitor candidate has been detected in archival images at wavelengths 
($\geq0.6~\mu{\rm m}$) where RSGs typically emit profusely. This object is distinctly variable in the infrared (IR). We characterize the variability using pre-explosion mid-IR (3.6 and $4.5~\mu{\rm m}$) {\sl Spitzer} and ground-based near-IR ($JHK_{s}$) archival data jointly covering 19~yr. The IR light curves exhibit significant variability with RMS amplitudes in the range of $0.2\mbox{--}0.4$~mag, increasing with decreasing wavelength. From a robust period analysis of the more densely sampled {\sl Spitzer} data, we measure a period of $1091\pm71$~d. 
We demonstrate using Gaussian Process modeling that this periodicity is also present in the near-IR light curves, thus indicating a common physical origin, which is likely pulsational instability. We use a period-luminosity relation for RSGs to derive a value of $M_{K}=-11.58\pm0.31$~mag. Assuming a late M spectral type, 
this corresponds to $\log(L/L_{\odot})=5.27\pm0.12$ at $T_{\rm eff}=3200$~K and to $\log(L/L_{\odot})=5.37\pm0.12$ at $T_{\rm eff}=3500$~K. 
This gives an independent estimate of the progenitor's luminosity, unaffected by uncertainties in extinction and distance. Assuming the progenitor candidate underwent enhanced dust-driven mass-loss during the time of these archival observations, and using an empirical period-luminosity-based mass-loss prescription, we obtain a mass-loss rate of around $(2\mbox{--}4)\times10^{-4}~{\rm M}_{\odot}~{\rm yr}^{-1}$.  
Comparing the above luminosity with stellar evolution models, we infer an initial mass for the progenitor candidate of $20\pm4~M_{\odot}$, making this one of the most massive progenitors for a Type II SN detected to-date.      

\end{abstract}


\section{Introduction}\label{sec:intro}
Supernova (SN) 2023ixf, which exploded in the nearby galaxy Messier 101 (M101; NGC 5457), was discovered by \citet{Itagaki} on 2023 May 19 (UT). It led to a flurry of observations given its proximity.  
Based on the first spectrum obtained on the night of discovery, SN~2023ixf was classified as a Type~II~SN by \citet{Perley_2023ixf}. During the early days of its evolution, it showed a strong blue continuum and notable flash-ionization features of the hydrogen Balmer-series, He~{\sc ii}, N~{\sc iv}, and C~{\sc iv}, which disappeared after a few days. These features may indicate the presence of He/N-rich circumstellar material (CSM) in the close vicinity of the explosion site \citep{Yamanaka_2023ixf,Jacobson-Galan_2023ixf,Smith_2023ixf}. The latter study also noted a lack of narrow blue-shifted absorption in the early spectra, which they argued to be a sign of strong asphericity of the CSM.

A further sign of ongoing CSM interaction in SN~2023ixf is the early detection of the object in hard X-rays (up to 20 keV) with NuSTAR \citep[at +4 and +11 days,][]{Grefenstette_2023ixf}, with the Mikhail Pavlinsky ART-XC telescope onboard the SRG observatory \citep[at +6 and +9 days,][]{Mereminskiy_2023ixf_ATel}, and with {\it Chandra} \citep[at +13 days,][]{Chandra_2023ixf_ATel}. At the same time, no evidence has been found for statistically significant emission in either the sub-millimeter \citep{Berger_2023ixf} or radio \citep[at 10~GHz,][]{Matthews_2023ixf_TNS} wavelengths within a few days after explosion. However, as \citet{Grefenstette_2023ixf} noted, radio emission may be suppressed by the high optical depth of the CSM at these early times and, if the assumed density profile were to remain the same at larger radii, then the SN may be expected to become radio/mm-bright a few weeks/months after explosion. Indeed, its radio emission was detected on a second epoch of observation with the VLA around +29~days by \cite{Matthews-2023}. This is consistent with the radio behavior of other SNe~II \citep[e.g.,][]{Weiler2010}.

Being one of the closest core-collapse SNe in several decades and located in a well-studied galaxy, significant efforts have been made to identify and analyze the progenitor of SN~2023ixf. First we identified a clear point source in archival images of the {\it Spitzer Space Telescope} (hereafter {\sl Spitzer}) at the absolute 
coordinates reported for the SN \citep[$\alpha_{\rm J2000}$=14:03:38.56, $\delta_{\rm J2000}$=54:18:42.02;][]{ZTF_2023ixf} and carried out a preliminary analysis of the pre-explosion infrared (IR) variability (\citealt{Szalai2023}; see details below). Archival {\it HST} images covering various wavelengths have also been quickly analyzed by us \citep{Soraisam-2023} and by \citet{Pledger-2023}. Both teams identified a clear source on the F814W image, but a counterpart was not detected in the bluer bands. This finding, together with the presence of a bright mid-IR source in pre-explosion {\sl Spitzer} images, already indicated that the identified progenitor candidate must be a luminous, dusty object (probably a red supergiant, RSG).

In several further studies, sets of either pre- or post-explosion data have been used for more detailed analyses of the assumed progenitor. \citet{Neustadt-2023} used the first reported {\sl Spitzer} and {\it HST} data to estimate the luminosity of the source; moreover, they also analyzed a roughly $15$~year-long dataset obtained with the Large Binocular Telescope (LBT) and found no evidence for pre-explosion outbursts in the optical. 
\citet{Kilpatrick-2023} and \citet{Jencson_2023ixf} have analyzed in detail the archival {\sl Spitzer}, {\sl HST}, and ground-based near-IR datasets of the object to obtain estimates of the luminosity, temperature, mass-loss rate, and initial mass of the assumed progenitor. We discuss their results in more detail later in this work. Additionally, \citet{Hosseinzadeh_2023ixf} analyzed the early-time ultraviolet (UV) and optical light curves of SN~2023ixf and determined a radius of around $410~R_{\odot}$ for the exploding star (consistent with an RSG).

Various studies have shown that RSGs exhibit semi-regular variations in their optical light, which are considered to be pulsations driven by the $\kappa$ mechanism in their hydrogen ionization zone coupled with convection \citep[e.g.,][]{Stothers-1969, Heger-1997, Guo-2002}. A period-luminosity relation has also been established for these objects \citep[e.g.,][]{Kiss-2006, Yang-LMC, Soraisam-2018, Chatys-2019}. 
In particular, \cite{Soraisam-2018} found a positive correlation between the observed optical variability amplitudes of the RSGs in M31 and their luminosity. A similar trend in the mid-IR was also noted by \citet{Yang_2018}. If the progenitor candidate for SN~2023ixf has a high luminosity, then its variability amplitude in the optical will be large, which could, in turn, modulate the emission from its CSM in the mid-IR. 

In this paper, we analyze archival {\sl Spitzer} and ground-based near-IR data, jointly covering a duration of 19~yr prior to the SN explosion, in order to get a detailed picture of the IR variability (and, thus, of the true nature) of the assumed progenitor of SN~2023ixf. 
A companion paper will present the overall properties of this candidate (Van~Dyk et al.~in prep.; Paper~II hereafter). We adopt the explosion epoch of MJD 60082.75 \citep[from][]{Hosseinzadeh_2023ixf}. We describe the datasets used in Sect.~\ref{sec:archive} and present our analysis results in Sect.~\ref{sec:analysis}. We discuss these results in Sect.~\ref{sec:dis} and conclude in Sect.~\ref{sec:conclude}.

\section{Available Archival Data}\label{sec:archive}

\subsection{{\sl Spitzer} IRAC}\label{spitzer}

\begin{figure*}[h]
\centering
\includegraphics[width=\columnwidth]{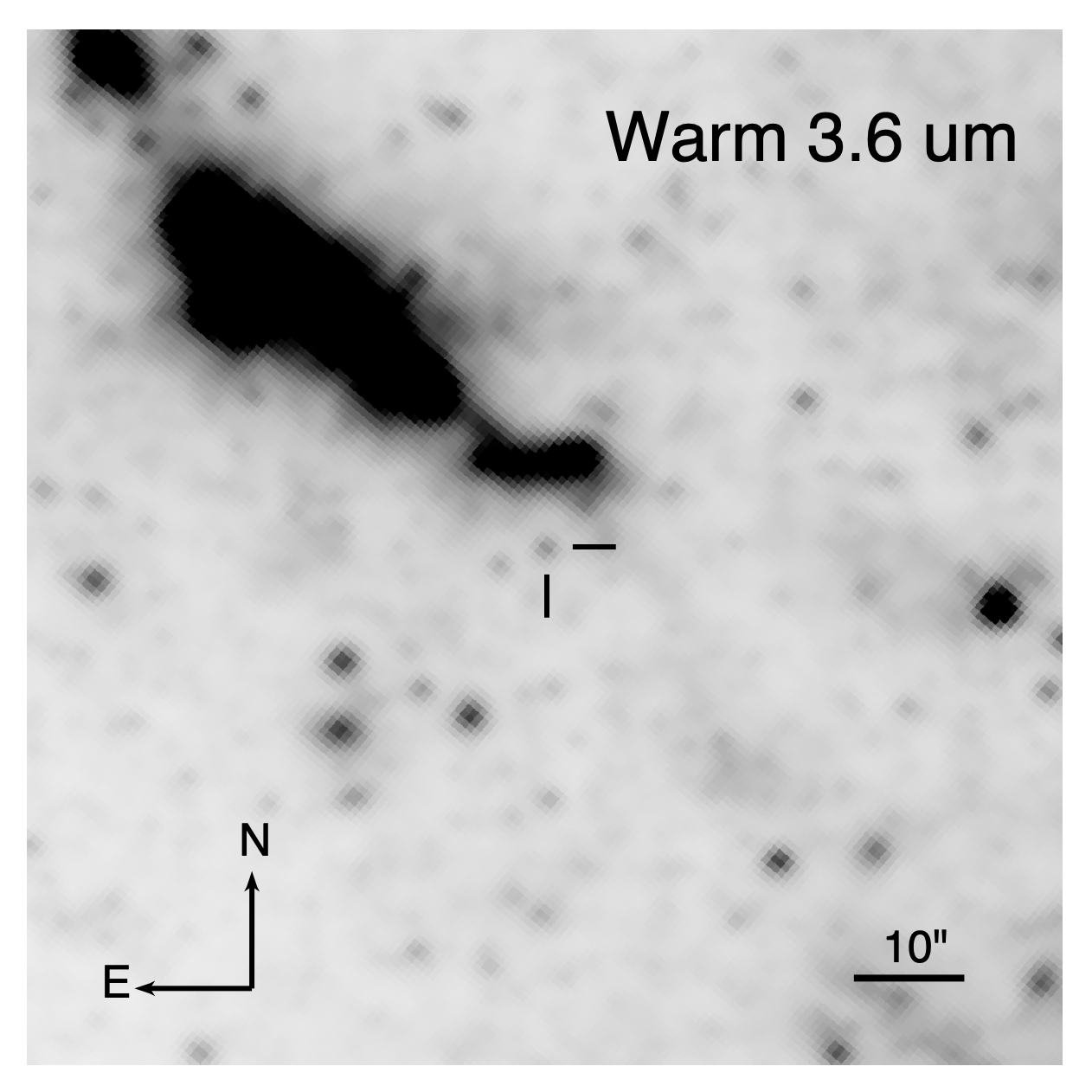}\hfil
\includegraphics[width=\columnwidth]{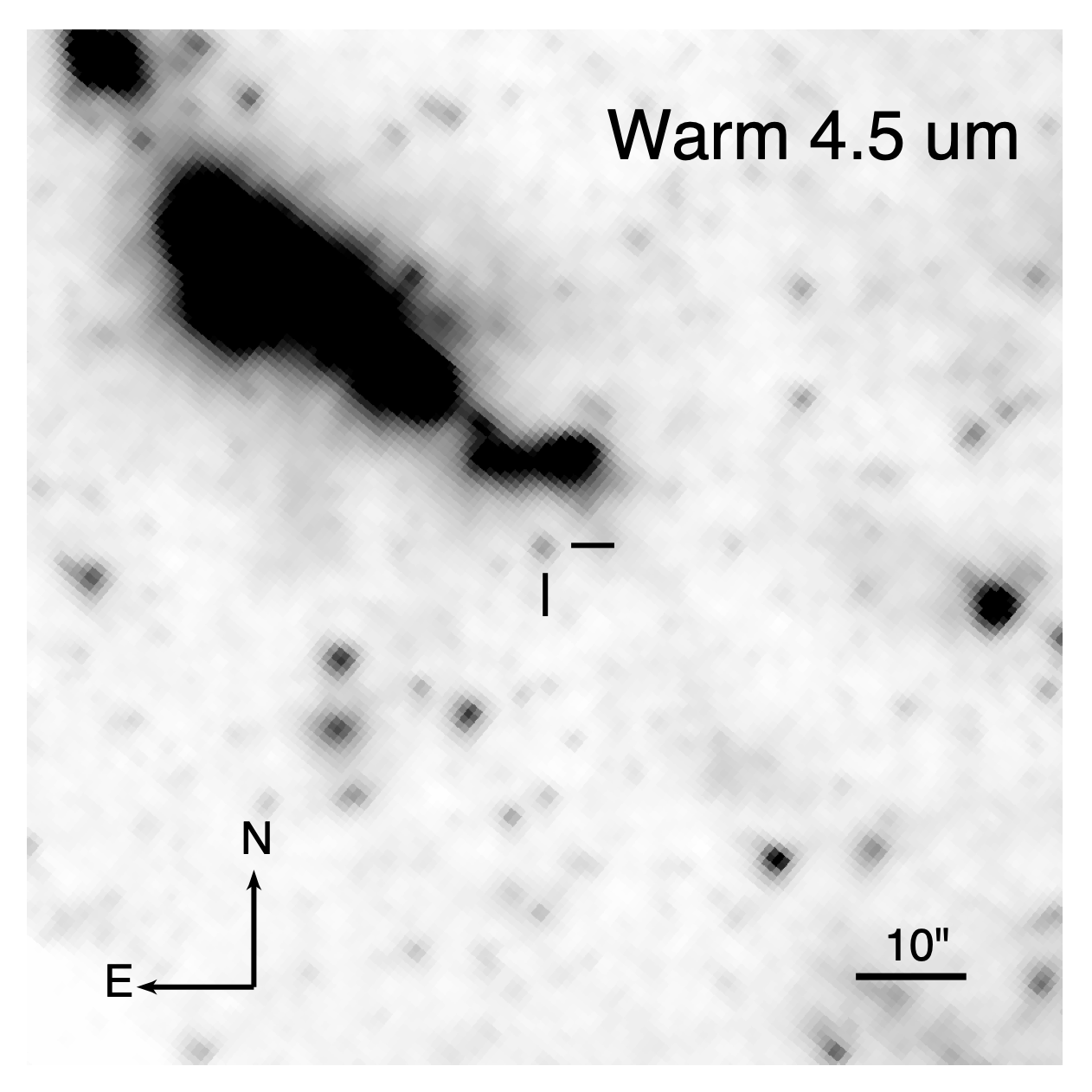}
\caption{Pre-explosion {\sl Spitzer\/} Warm Mission mosaics at Channels 1 and 2 (3.6 and 4.5 $\mu$m, respectively). A source can be clearly seen at the absolute position of SN~2023ixf in both channels, which we identify as its progenitor candidate in the mid-IR. 
}
\label{fig:spitzer_images}
\end{figure*}

Since M101 was a target of several programs during the {\sl Spitzer\/} mission, we checked the {\sl Spitzer\/} Heritage Archive\footnote{\href{https://sha.ipac.caltech.edu/}{https://sha.ipac.caltech.edu/}} (SHA; at the NASA/IPAC Infrared Science Archive, or IRSA) for available data, in order to follow the photometric evolution of the progenitor on the archival mid-infrared (mid-IR) images. A single-epoch dataset exists during the {\sl Spitzer\/} Cryogenic Mission (2004 March 08, PID 60 [GORDON-M101], PI: G.~Rieke), while the field of the SN was imaged on many additional epochs between 2012 and 2019 (mostly during the SPIRITS program, \citealt{Kasliwal2017}; PIDs 10136, 11063, 13053, 14089; PI: M.~Kasliwal). A point source is clearly detectable on 3.6 and 4.5~$\mu{\rm m}$ (Channels 1 and 2, respectively; see Fig.~\ref{fig:spitzer_images}) {\sl Spitzer}/IRAC images at the position of the SN 
at all epochs between 2004 and 2019, while there is no detection on 5.8 and 8.0~$\mu{\rm m}$ images obtained in 2004. We considered this source to be the progenitor candidate and reported on our preliminary analysis in \citet{Szalai2023}, whereby we determined the mid-IR fluxes of the source by applying simple aperture photometry on the 3.6 and 4.5~$\mu$m {\sl Spitzer}/IRAC post-basic calibrated (PBCD) images. The measurements showed some possible flux changes, though below the photometric uncertainties, in the preceding $19$~yr before the explosion. If this variability is real, it could be an indication of possible internal pulsational instabilities and pre-explosion mass-loss processes occurring in the star years prior to explosion. Hence, we carried out a more careful analysis of the archival {\sl Spitzer}/IRAC data of the object, concurrent with that of \citet{Kilpatrick-2023} and \citet{Jencson_2023ixf}.

Rather than continuing with the PBCD mosaics, we obtained the artifact-corrected cBCD frames (along with the corresponding uncertainty and mask frames) for each epoch. The Cryogenic and Warm data were treated separately. We further selected for each epoch only the frames covering the location of the progenitor candidate. 
We then performed overlap correction and mosaicking of these frames with {\tt MOPEX} \citep{Makovoz2005a}. 
We invoked the task {\tt APEX} User List Multiframe, with the absolute SN position as input, to extract the flux via fitting of the Point Response Function (PRF) for each epoch. A visual spot-check of the residual mosaics after the PRF fitting, generated with {\tt APEX QA}, revealed a clean subtraction of the PRF model in each case. We list the {\sl Spitzer\/} flux densities, along with their formal uncertainties (based on SNR), in Table~\ref{tab:spitzer}. We note that the absolute photometric calibration of the entire {\sl Spitzer\/} mission is better than 3\% \citep{Carey2012}.

To validate our PRF photometry results and to provide more realistic estimates of the uncertainties, we injected an artificial star into the cBCD frame for each epoch, with a flux equal to that estimated for the progenitor candidate, at a distance of $8{\farcs}4$ east-southeast of the candidate, in an empty region with a qualitatively similar background as the candidate, using {\tt APEX QA}. The overlap, mosaic, and PRF estimation steps were then repeated, with both the candidate and the artificial star extracted. We found for both bands that the flux of the artificial star was on average ca.\ $3\ \mu$Jy, or $11$\%, brighter than the estimate for the candidate. A similar spot check of the residual mosaics, again, indicated clean and complete PRF fitting for both the candidate progenitor and the artificial star. We have therefore included this slight flux excess into the uncertainties for the measurements in each epoch.

\begin{figure*}
\centering
\includegraphics[width=150mm]{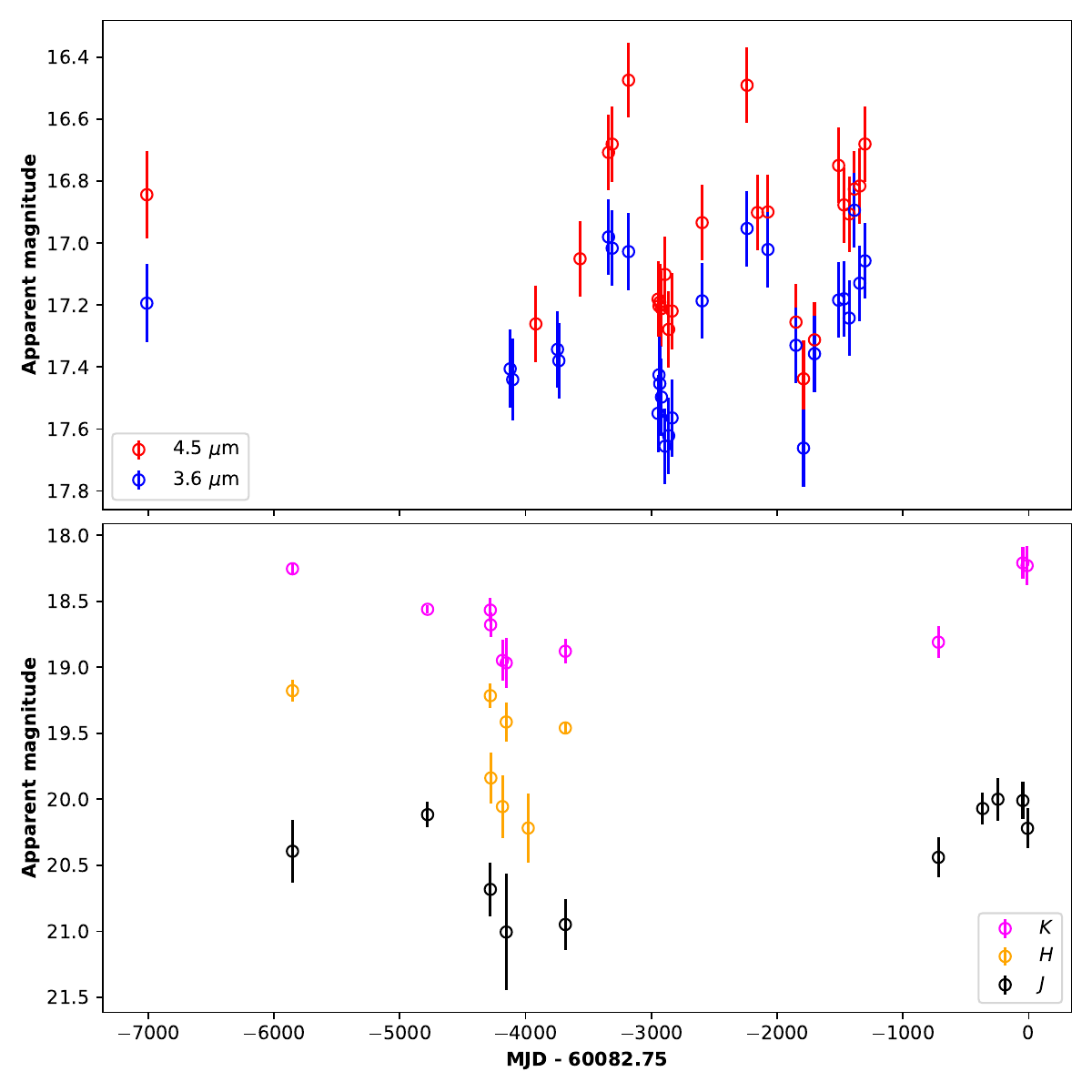}
\caption{Light curves of the progenitor candidate in the mid-IR based on {\sl Spitzer} data (upper panel) and the near-IR based on UKIRT/WFCAM and Gemini/NIRI data, along with MMT/MMIRS data reported by \citet{Jencson_2023ixf} (lower panel). The x-axis origin corresponds to the explosion epoch of SN~2023ixf \citep{Hosseinzadeh_2023ixf}.  
}
\label{fig:spitzer_lc}
\end{figure*}

To convert the flux densities into Vega magnitudes\footnote{All magnitudes reported are in the Vega system.}, we adopted the IRAC zeropoints, $272.2\pm4.1$~Jy and $178.7\pm2.6$~Jy at the nominal channel wavelengths of 3.544 and $4.487\ \mu$m, respectively. The light curves for these two channels are shown in the top panel of Fig.~\ref{fig:spitzer_lc}. Significant variability can be clearly seen. We measure a root-mean-square (RMS) amplitude of 0.23 and 0.26~mag for Channels 1 and 2, respectively, in good agreement with the average variation found by \citet{Reiter_2015} for Galactic long-period variables (LPVs).

\citet{Kilpatrick-2023} have also used this same {\sl Spitzer\/} data set and noted the variability, which is similar to our value of 0.7 mag peak-to peak. Their measured fluxes are, however, $\sim 5\mbox{--}10~\mu{\rm Jy}$ lower than ours, which is likely due to differences in the photometry methods. 
They have performed sanity checks to ascertain that the variability is intrinsic to the star and not due to instrumental or spacecraft-related systematic effects. Hence, we proceed below considering the variability to be indeed intrinsic to the star. Additionally, in the archival {\sl HST} ACS/WFC F814W pre-explosion image, there is possibly an indication of blending of the progenitor star candidate with another (fainter) source located $0{\farcs}1$ from it \citep[e.g.,][]{Pledger-2023, Kilpatrick-2023}. However, the observed mid-IR variability is likely dominated by a single source (see Sect.~\ref{sec:analysis}).      

\citet{Jencson_2023ixf} also analyzed the archival {\sl Spitzer} data of the assumed progenitor in detail. They identified the same source as we did, 
carried out PSF photometry on the Cryogenic Mission data and, following that, performed image subtraction and aperture photometry on the Warm Mission data using the Cryogenic mosaic images as templates. 
Results of their $4.5~\mu{\rm m}$ photometry agree well with ours; however, for the $3.6~\mu{\rm m}$ photometry, there is an offset similar to that of \citet{Kilpatrick-2023}. We note that the discrepancies likely arise from differences in the photometric techniques. The general light curve variability, however, is the same between this work and that of \citet{Jencson_2023ixf} and \citet{Kilpatrick-2023}. Thus, the offsets in the photometry noted above do not affect the main results of our paper, which are based on the variability in the light curves, in particular their period analysis (cf.~Sect.~\ref{sec:analysis}). We find a similar period using the $3.6~\mu{\rm m}$ light curve from \citet{Jencson_2023ixf} as well as the one from \citet{Kilpatrick-2023}.

\subsection{Gemini NIRI} \label{niri}

\begin{figure*}[ht]
\centering
\includegraphics[width=80mm]{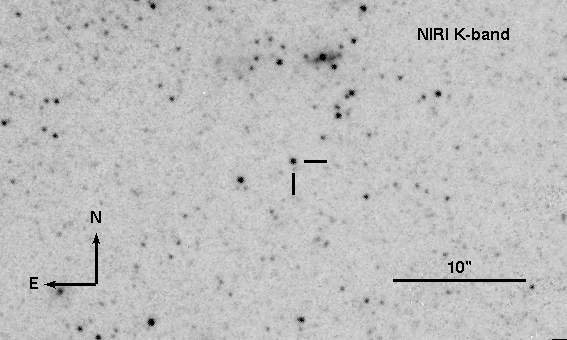}\hfil\includegraphics[width=80mm]{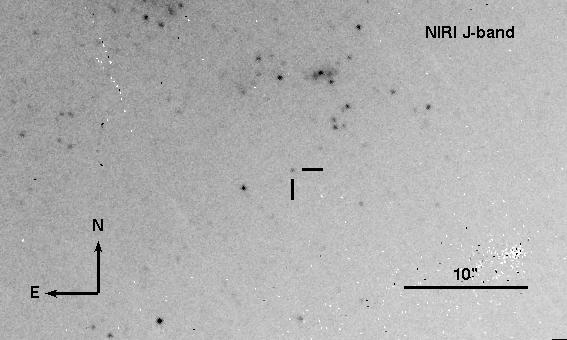}\\
\vspace{0.5cm}
\includegraphics[width=80mm]{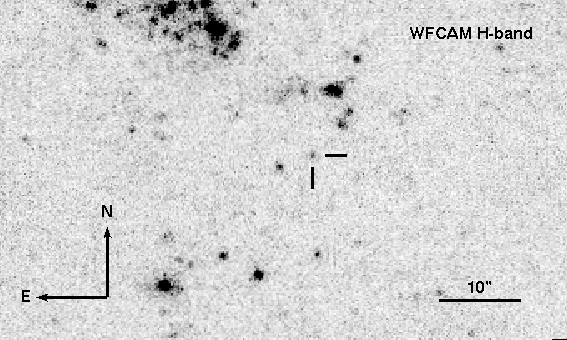} 
\caption{Cutouts around the progenitor candidate for SN~2023ixf from archival NIRI $K$-band (top left), $J$-band (top right), and UKIRT/WFCAM $H$-band (bottom) images.}
\label{fig:niri_Kcutout}    
\end{figure*}

We searched the Gemini Science Archive \citep{gem_archive} for any available pre-explosion data at the position of SN~2023ixf. There are observations taken with the Near-IR Imager (NIRI; \citealt{niri}) on 2010 April 18 (UT) for the program GN-2010A-Q-27 (PI: G.~Bosch) to study the HII region NGC~5461 located near to SN~2023ixf, thus providing serendipitous pre-explosion imaging for the latter. This program used two filters---$K$-continuum and narrow-band Br(gamma) with central wavelengths $2.0975~\mu{\rm m}$ and $2.1686~\mu{\rm m}$, respectively. There are 26 exposures of Br(gamma) and 51 exposures of $K$-continuum, each of 50~sec. In addition, there are two $J$-band acquisition images, each with an exposure time of 30~sec. In the following, we do not use the narrow-band data as the wavelength is already covered by the $K$-continuum band. 

We reduce these data with Gemini's DRAGONS software \citep{dragons} using its recipe for NIRI imaging data, which performs dark subtraction, masking of bad pixels, flat fielding, sky subtraction, image registration, and stacking. The image quality for one of the $J$-band acquisition images is rather poor, so we do not include that exposure. For this filter, we thus have only one image so no sky subtraction has been performed for it by the DRAGONS processing. This does not affect our result, since the background is subtracted during the downstream processing for photometry.   

A point source at the position of SN~2023ixf is clearly detected in both the $K$-continuum and $J$-band images as shown in Fig.~\ref{fig:niri_Kcutout}. We perform PSF photometry using the Photutils package of Python \citep{photutils} and apply the aperture correction resulting from a curve-of-growth analysis for each image. Since there are only two 2MASS stars in the NIRI image, we calibrate the final photometry using five isolated stars in the images by comparing their instrumental magnitudes to corresponding calibrated magnitudes measured from the UKIRT/WFCAM data discussed below (Sect.~\ref{ukirt}). We thus obtain $J=20.12\pm0.10$~mag and $K=18.56\pm0.03$~mag for the progenitor (cf.~Table~\ref{tab:nir}).

We note that \citealt{Kilpatrick-2023} have also used the $K$-continuum data from this Gemini program (but not the $J$-band data), and calibrated their result against the NOIRLab Extremely Wide Field Infrared Imager (NEWFIRM) $K$ data observed around the same time (2010 June$\mbox{--}$July). They measured a value of $K=20.72\pm0.08$~AB mag from the NIRI data for the progenitor candidate, which is $18.87\pm0.08$ converted to Vega mag; they obtained a similar value for the NEWFIRM $K$ data. Their result is about 0.3~mag fainter than ours. This could perhaps be due to differences in the data processing. Since the NEWFIRM data do not add any additional information to that already contained in the Gemini data, we do not consider it further in our analysis.

\subsection{UKIRT/WFCAM} \label{ukirt}

We also searched the archival data gathered by the near-infrared (near-IR) Wide Field Camera (WFCAM; \citealt{wfcam}) on the 3.8~m United Kingdom Infrared Telescope (UKIRT). Science-quality reduction of the WFCAM data is performed by the automated WFCAM/VISTA pipeline at the Cambridge
Astronomical Survey Unit \citep{ukirt_pipeline} and the data products are then ingested into the WFCAM Science Archive (WSA; \citealt{ukirt_archive}). Astrometric and photometric calibrations, both tied to 2MASS \citep{2MASS}, are carried out as part of the pipeline processing. In particular, the pipeline derives a final photometric calibration in the WFCAM system (see \citealt{Hodgkin-2009} for details).

We find 41 images for each of $J$-, $H$-, and $K$-band between 2007 and 2013, with exposure times ranging from 1 to 20~sec. The project IDs for these observations are U/11B/K1, U/11B/K2, U/11B/K4, U/12A/KASI1, U/13A/K1, and U/SERV/1745. Since many of the images are shallow and taken close in time, we decide to stack all the images observed in the same month in a given year. The stacking is not expected to affect our analysis since the timescale of variability for RSGs is typically $100$~d or greater \citep[e.g.,][]{Jurcevic-2000, Kiss-2006, Yang-LMC, Yang-SMC, Wasatonic-2015, Soraisam-2018, Ren-2019, Chatys-2019}. A source can be clearly seen at the position of SN~2023ixf in many of the images (e.g., the $H$-band image taken on 2007 May 10 (UT), shown in Fig.~\ref{fig:niri_Kcutout}, bottom panel), which we consider as the progenitor candidate in the near-IR.  

We perform (forced) PSF photometry on the WFCAM data at the SN position obtained from its first detection{\footnote{https://antares.noirlab.edu/loci/ANT2023l4lgj6bhp4rt \citep{Antares}}} at $g=15.90\pm0.03$ on 2023 May 19 (UT) by the ZTF survey \citep{Bellm-2019}. It is to be noted that the astrometric accuracy for ZTF (tied to Gaia~DR1) is better than a few tens of milliarcsec for bright sources (see \citealt{Masci-2019}), while that for WFCAM is $0{\farcs}1$ \citep{Hodgkin-2009}. For all three filters, the offset between the PSF-fitted position of the progenitor candidate and that of the SN is less than 1~pixel (i.e., $0{\farcs}2$) of the WFCAM image. Similar to that of NIRI data, we then determine the aperture correction for the PSF photometry from a curve-of-growth analysis. The total instrumental magnitude is then calibrated using the available zero-point{\footnote{For the WFCAM image stacks, we recompute the zero-points taking one of the exposures whose zero-point is available as the reference.}} of the image. The resulting near-IR measurements are given in Table~\ref{tab:nir} and the light curves are shown in Fig.~\ref{fig:spitzer_lc}.

We also plot the most recent $J$ and $K$ photometry of the progenitor candidate obtained by \citet{Jencson_2023ixf} using MMT/MMIRS. These are the data points within 1000~d of the explosion in Fig.~\ref{fig:spitzer_lc}, lower panel. From the sparse near-IR light curves, we obtain RMS amplitudes of 0.36, 0.38, and 0.28~mag, respectively, for the $J$-, $H$-, and $K$-band.

\section{Analysis}\label{sec:analysis}

\begin{figure}
\centering
\includegraphics[width=80mm]{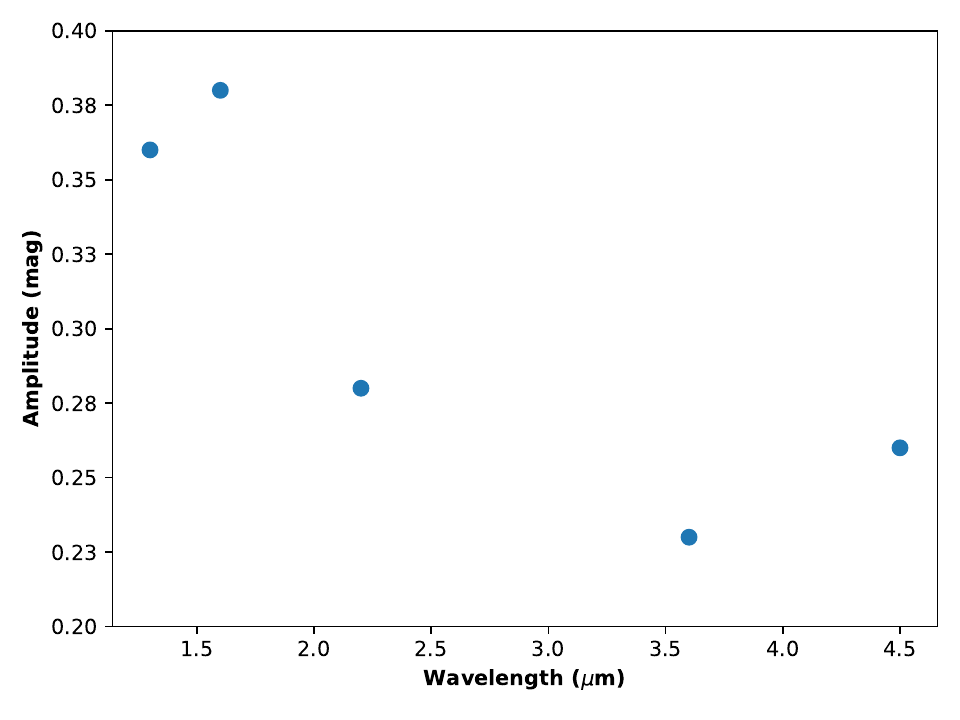}
\caption{Variation of light curve RMS-amplitude of the progenitor candidate as a function of wavelength.}
\label{fig:amp_wave}
\end{figure}

\begin{figure*}
\centering
\includegraphics[width=80mm]{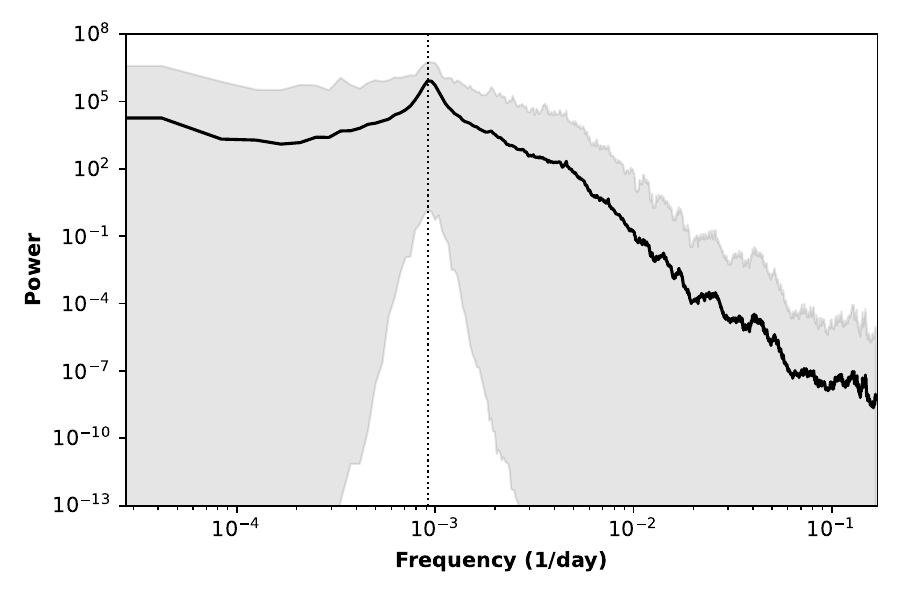}\hfil\includegraphics[width=80mm]{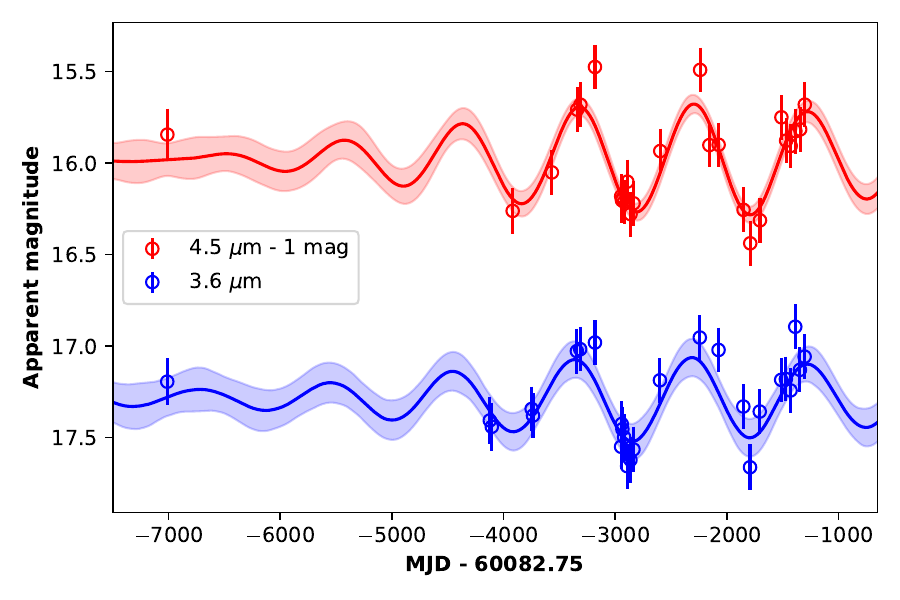} 
\caption{{\it Left}: Reconstructed power spectrum of the {\sl Spitzer} $3.6~\mu{\rm m}$ light curve with its 99 percent confidence interval highlighted by the gray shaded region. The gray dotted line marks the location of the identified period. {\it Right}: Reconstructed light curves using the power spectrum on the left for the {\sl Spitzer} 3.6 and $4.5~\mu{\rm m}$ data shown by the blue and red curves, respectively. The shaded color regions mark their $1\sigma$ confidence intervals.}
\label{fig:spit_recon}
\end{figure*}

In Fig.~\ref{fig:amp_wave}, we show the RMS amplitudes from the mid-IR and near-IR light curves as a function of wavelength. There is a clear indication of an increase in variability amplitude with decreasing wavelength, which is in line with expectations for RSGs and other LPVs. For example, \citet{Wasatonic-2015} analyzed 17~yr of $V$-band and narrow- to intermediate-band near-IR high-cadence time-series photometry of the Galactic RSG TV~Gem and found the peak-to-peak amplitude to increase from 0.3~mag in the near-IR to over $1$~mag in the $V$-band. Similarly, \citet{Wood-1983} measured a $K$-band amplitude of $\lesssim0.25$~mag for RSGs in the Large and Small Magellanic Clouds, while \citet{Yang_2018} found an amplitude $\lesssim0.3$~mag in the mid-IR (WISE1- and WISE2-band) for the sample of RSGs in the Large Magellanic Cloud that they studied. Also, \citet{Boyer_2015} cross-identified variable sources from the DUSTiNGS survey with LPVs that have been observed in the optical and near-IR and found the variability amplitude to increase with decreasing wavelength.

We perform a period analysis using the {\sl Spitzer} light curves. As described in previous work by various authors, e.g., \citet{Kiss-2006}, \citet{Soraisam-2018}, light curves of RSGs present complex morphologies, characterized by the superposition of a strong red-noise component and one or multiple distinct frequencies. For analysis of such semi-regular variables, conventional methods like Lomb-Scargle periodogram may be 
inadequate \citep{VanderPlas-2018}. We therefore utilize the same Bayesian model used by \citet{Soraisam-2018} for the period analysis of optical light curves of RSGs in M31. Specifically, we describe the $3.6~\mu{\rm m}$ light curve as a Gaussian Process using the \texttt{NIFTy} software package \citep{Selig-2013, nifty-3, nifty-5}. With the latter we simultaneously reconstruct the underlying signal and its unknown power spectrum from the observed light curve data and reported uncertainties (see Sect.~2.4 of \citealt{Soraisam-2018} for details), by sampling an approximate representation of the joint posterior distribution of light curve and power spectrum. The results are shown in Fig.~\ref{fig:spit_recon}.

We use the recovered power spectrum from the $3.6~\mu{\rm m}$ data to model the $4.5~\mu{\rm m}$ light curve, again as a Gaussian Process but this time with a fixed power spectrum{\footnote{It is to be noted that the key component of a Gaussian process, i.e., its covariance matrix, is determined by the power spectrum, assuming stationarity of the signal.}}, which is highlighted in the right panel of Fig.~\ref{fig:spit_recon}. As can be clearly seen, the reconstructed light curve fits the data well, which implies that the periodicity seen in both the {\sl Spitzer} channels is the same, likely due to both of them having the same physical origin. We obtain similar results when executing this process in reverse, i.e., using the $4.5~\mu{\rm m}$ data for inferring the power spectrum and using it to model the $3.6~\mu{\rm m}$ light curve.

For each of the posterior power spectrum samples, we determine the frequency corresponding to the power maximum ({\em peak frequency}). We then use the median of these peak frequencies to define the period for the {\sl Spitzer} $3.6~\mu{\rm m}$ light curve, obtaining $1091\pm71$~d 
(marked by the dotted line in Fig.~\ref{fig:spit_recon}, left panel), where the uncertainty is derived from the width of the distribution of peak frequency values. 

As mentioned in Sect.~\ref{spitzer}, there is possibly a fainter source located close to the progenitor candidate in the optical {\sl HST} F814W pre-explosion image. This opens up the question of possible blending of sources (including those not visible in the optical) for the mid-IR data, given their poorer spatial resolution as compared to {\sl HST}. We argue that it is unlikely that all the blended sources vary with exactly the same dominant frequency to give rise to the peak in the power spectrum in Fig.~\ref{fig:spit_recon}. The observed variability at the position of SN~2023ixf therefore most likely results from a single source, and we attribute that to the RSG  progenitor candidate.

\section{Discussion}\label{sec:dis}

\subsection{Pulsational instabilities?}\label{pulsation}

\begin{figure*}
\centering
\includegraphics[width=\textwidth]{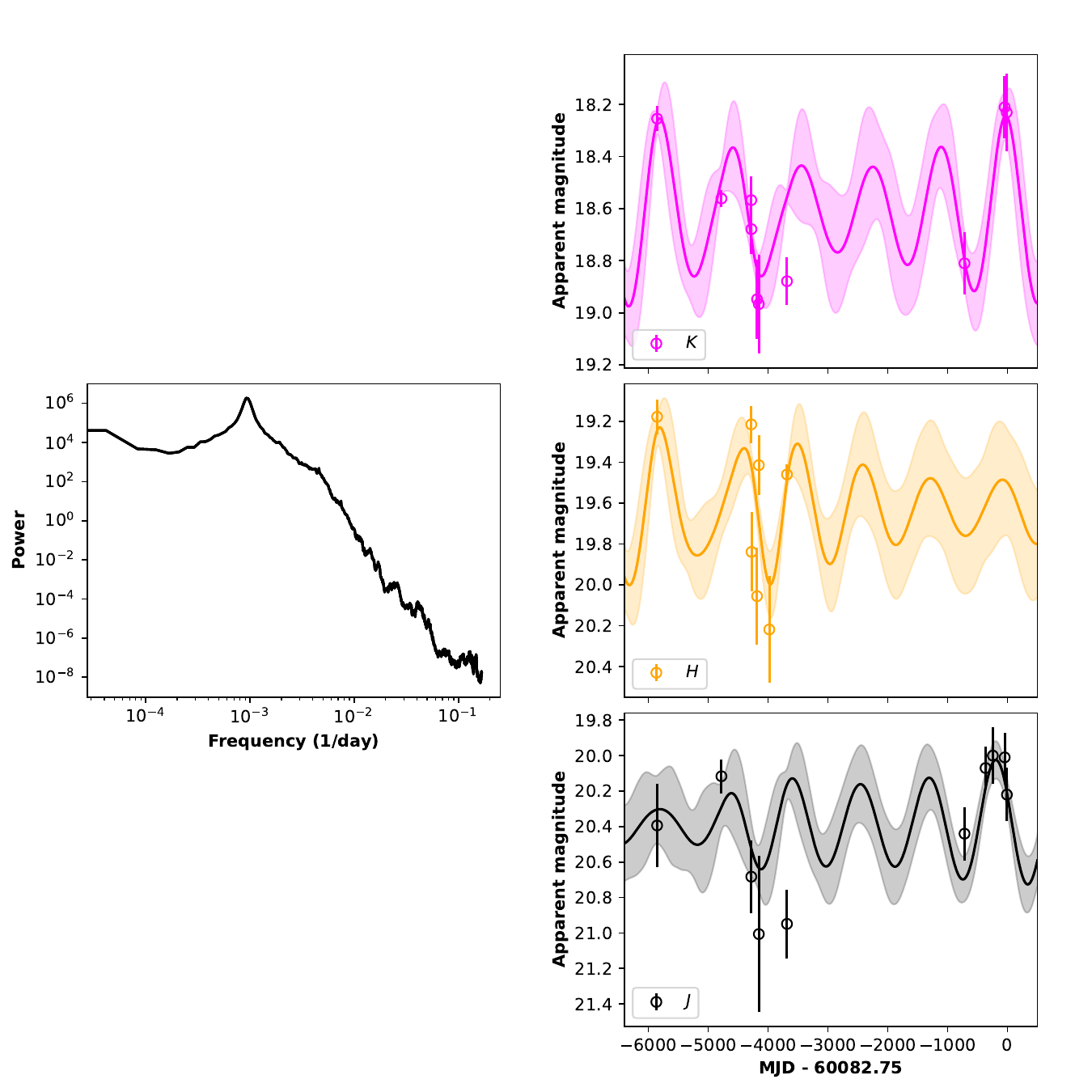}
\caption{Near-IR light curve models for the progenitor candidate of SN~2023ixf constructed using the power spectrum on the left, which is recovered from the {\sl Spitzer} data (Fig.~\ref{fig:spit_recon} left) scaled by a factor 2.2 to account for the larger variability amplitude in the near-IR (Fig.~\ref{fig:amp_wave}). The shaded regions indicate the $1\sigma$ confidence intervals for the models. Despite the scarce data points, the light curves look reasonable as compared to those built using various modified versions of the power spectrum shown in Figs.~\ref{fig:bad_2}, \ref{fig:bad_3}, \ref{fig:bad_4}.}
\label{fig:nir_recon}
\end{figure*}

The near-IR emission of RSGs is dominated by light from the stellar photosphere. 
To ascertain whether the same periodicity can also explain the near-IR light curves, we model the latter using the power spectrum of the {\sl Spitzer} $3.6~\mu{\rm m}$ light curve obtained above. Given the slightly larger amplitude of the near-IR light curves as compared to the mid-IR (Fig.~\ref{fig:amp_wave}), we scale this power spectrum by a factor 2.2, which is the average of the ratios of the squares of the amplitudes in the near-IR to the mid-IR{\footnote{There is no dramatic difference if we use individual scaling factors for the $J$, $H$, and $K$ bands.}}. The resulting models are shown in the right panel of Fig.~\ref{fig:nir_recon}. As is evident from the plots, the models agree well with the data, implying the presence of the same periodicity in the near-IR data and, in turn, indicating again a common driver for the variability observed in both the near-IR and mid-IR regimes. We further demonstrate in Appendix~\ref{app} that similar frequencies are needed to explain the near-IR data, by showing reconstructions performed with power spectra where the peak around a period of 1091~d is absent.  

Indeed, near-IR variability in RSGs driven by pulsation has been established for decades \citep[e.g.,][]{Wood-1983}. As discussed in Sect.~\ref{sec:analysis}, the near-IR amplitudes for the progenitor candidate are also in agreement with those of other known pulsating RSGs and therefore we conclude that pulsational instabilities in the progenitor star are likely responsible for its observed variability. 

As shown by \citet[][see their Fig.~3]{Davies-2022}, mid-IR emission from the pre-SN outburst models stays almost constant, while the optical and near-IR bands may be significantly affected and thereby provide the best diagnostic for such outbursts. 
The MMT/MMIRS $J$ and $K$ band observations taken less than two weeks prior to the SN explosion appear to follow the general variability trend in the near-IR going back $16$~yr based on the WFCAM and NIRI data (Fig.~\ref{fig:nir_recon}), thus supporting the results obtained by other groups that there was likely no pre-SN outburst \citep{Kilpatrick-2023, Neustadt-2023, Jencson_2023ixf, Dong-2023}. There are not enough deep optical observations of the progenitor prior to explosion to measure the variability that is typically present for RSGs in this wavelength regime. However, assuming its pulsational instability to be similar to that in common RSGs, it is likely to have a larger variability amplitude in the optical. For the RSG TV~Gem, \citet{Wasatonic-2015} measured the $V$-band amplitude to be a factor $4$ larger than that in the near-IR. Using the same scaling, we estimate the optical amplitude of the RSG progenitor candidate to be around 1.6~mag.

\subsection{Estimates of the progenitor's initial mass and mass-loss rate}\label{mass}

\begin{figure}
\centering
\includegraphics[width=80mm]{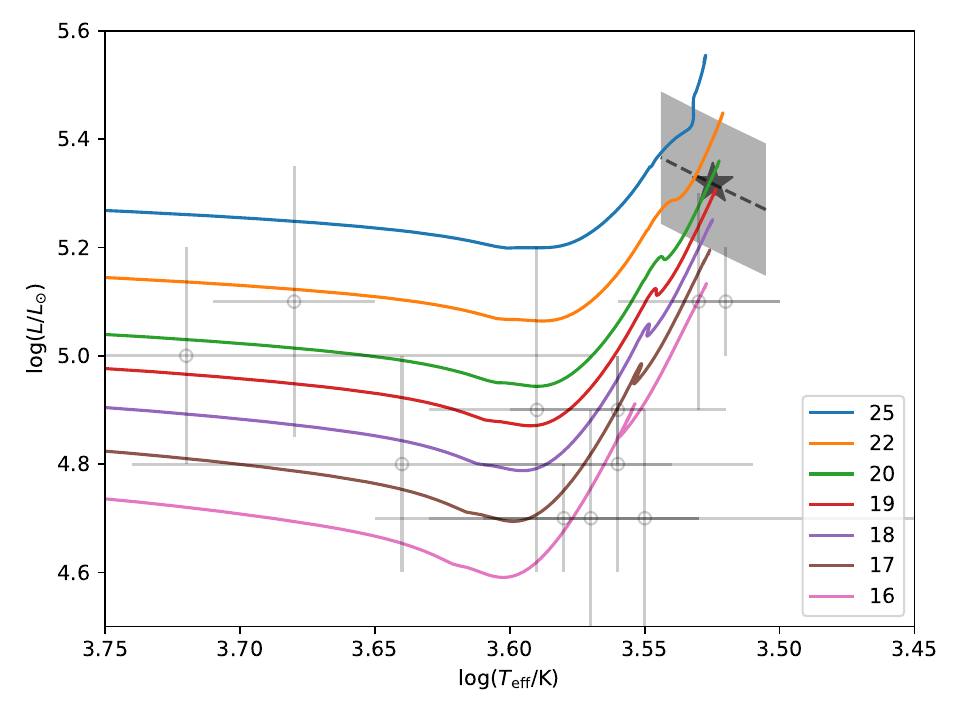}
\caption{HR diagram based on the BPASS single star evolutionary tracks. Different colors indicate different initial masses of the models as indicated in the legend (in ${\rm M}_\odot$). The gray shaded parallelogram around the $\bigstar$-symbol shows the possible location of the progenitor candidate of SN~2023ixf, assuming a late M-type supergiant with $T_{\rm eff}$ ranging from $3200\mbox{--}3500$~K (see text). The vertical width of the region corresponds to the uncertainty in $\log(L/L_{\odot})$ at a given temperature. The faint gray crosses mark the locations of cool and luminous ($\log(L/L_{\odot})>4.7$), directly detected progenitors of previous SNe-IIP/II-L from \citet{Smartt-2015} and \citet{SVD-2017}.}   
\label{fig:mass}
\end{figure}

We use the period of 1091~d derived above (Sect.~\ref{sec:analysis}) and the period-$M_{K}$ (PL) relation{\footnote{$M_{K}=(-3.38\pm0.27)\times\log(P)+(-1.32\pm0.75)$}} for RSGs from \citet{Soraisam-2018} to compute the absolute magnitude of the progenitor candidate in $K$-band. This method, which is fundamentally different from the widely used SED fitting, provides an estimate of the luminosity of the star that does not rely on using extinction information or the distance to the host galaxy. 
We thus obtain a value of $M_{K}=-11.58\pm0.31$~mag. \citet{Soraisam-2018} noted a dispersion of 0.29~mag around their period-$M_{K}$ relation. We have included this as well as the uncertainty in our derived period in the calculation of the final uncertainty of our $M_{K}$ estimate here. We note that we obtain values of $M_{K}$ that agree with our estimate within the uncertainties when using other published PL relations for RSGs \citep[e.g.,][]{Kiss-2006, Yang-SMC}. This may be expected since the PL relation for RSGs appears to be universal (cf.~Fig.~7 of \citealt{Soraisam-2018}).

\citet{Jencson_2023ixf} report a foreground-extinction-corrected and phase-weighted mean absolute magnitude of $M_{K}=-10.7$~mag directly computed from the observed light curve. This value is not corrected for CSM extinction, which is likely to be significant for the progenitor candidate. In addition, it will be sensitive to uncertainties in distance and interstellar extinction estimates. Nevertheless, it can be considered as a lower limit for the actual $M_{K}$, and thus, our brighter $M_{K}$ value is consistent with their result.

Taking into account this high value of $K$-band absolute brightness, as well as its near- and mid-IR colors ($[3.6]-[4.5]\approx0.3$~mag, $J-K\approx1.8$~mag), the assumed progenitor of SN~2023ixf seems to be among the most luminous RSG stars (for comparison, see the results of large-scale AGB/RSG surveys in nearby galaxies by \citealt{Boyer_2015,Grady-2020,Massey_2023}). In the SPIRITS sample analyzed by \citet{Bond_2022}, there are some objects (e.g., 15ahg, 15wt) with mid-IR light-curve periods and amplitudes, absolute magnitudes, and colors similar to that of the progenitor of SN~2023ixf; these objects are found to be mostly similar to dusty OH/IR stars by the authors. Such objects are known to undergo intensive mass-loss during the so-called {\em superwind} phase \citep{Iben-1983}. \citet{Jencson_2023ixf} have shown via comparison of the progenitor candidate's SED with superwind-based models from \citet{Davies-2022} that it likely underwent such a phase of enhanced mass-loss (rather than an outburst) over the final decades of its life.

In Paper~II, we perform SED-fitting of the progenitor candidate using the Grid of Red supergiant and Asymptotic giant branch star ModelS (GRAMS; \citealt{Sargent-2011, Srinivasan-2011}). 
In addition to uncertainties in distance and interstellar extinction toward the star, SED fitting is also sensitive to limitations in the models used, for example, coarse sampling of $T_{\rm eff}$ and optical depth parameters in the available GRAMS grid. We circumvented the latter issue in Paper~II by training an emulator (see Paper~II for details). However, we find that the $T_{\rm eff}$ value is not well constrained---the fit allows a large range from around 2400~K to 3700~K. 
Such a large range is consistent with the inferred relatively high dust obscuration of the progenitor candidate, which makes it difficult to predict the stellar parameters, in particular given that the sparse data in the mid-IR (covering only $3.6~\mu{\rm m}$ and $4.5~\mu{\rm m}$) only poorly constrain the dust content.

As noted above, the mid-IR properties of the progenitor candidate appear to be consistent with those of dusty OH/IR stars. 
Such stars typically have a late M spectral type---for example, M7 for IRC~-10414 (cf.~\citealt{Gvaramadze-2014}), M6 for NML~Cyg, M4--M9.5 for VX~Sgr, M3--M4 for S~Per \citep{Schuster-2006}, and M5 for WOH~G64 \citep{Levesque-2009}.   
Thus, to be completely independent of the SED-fitting results, we adopt the $T_{\rm eff}$ range of $3200\mbox{--}3500$~K spanned by M4 and later spectral type RSGs. We use the corresponding bolometric correction to the $K$-band ($BC_{K}$) determined by \citet{Levesque-2005} based on the MARCS stellar atmosphere models for cool stars to convert the absolute $K$-band magnitude to bolometric luminosity. The values of $BC_{K}$ range from 3.16 to 2.92~mag, decreasing with increasing temperature.

We thus obtain an absolute bolometric magnitude $M_{\rm bol}=-8.42\pm0.31$~mag  and a corresponding $\log(L/L_{\odot})=5.27\pm0.12$ at the lowest $T_{\rm eff}=3200$~K considered above, and $M_{\rm bol}=-8.66\pm0.31$~mag and corresponding $\log(L/L_{\odot})=5.37\pm0.12$ at $T_{\rm eff}=3500$~K. 
We note that the luminosity derived here from the PL relation is about $0.2\mbox{--}0.3$~dex higher than that obtained from SED fitting (Paper~II, \citealt{Jencson_2023ixf}).

The typical mass-loss rate for RSGs is in the range $10^{-7}\mbox{--}10^{-4}~{\rm M}_{\odot}~{\rm yr}^{-1}$ \citep[e.g.,][]{Mauron-2011}, while that for OH/IR stars is in the range $10^{-5}\mbox{--}10^{-3}~{\rm M}_{\odot}~{\rm yr}^{-1}$ \citep[e.g.,][]{Justtanont-2013}. \citet{Goldman-2017} derived an empirical mass-loss prescription for AGBs/RSGs in the superwind phase. They found that the mass-loss rate mainly correlates with the pulsation period and luminosity. We use their relation $\dot{M}=4.2\times10^{-11}L^{0.9}P^{0.75}r_{\rm gd}^{-0.03}$, where $\dot{M}$ is the mass-loss rate (in ${\rm M}_{\odot}~{\rm yr}^{-1}$), $L$, the luminosity (in $L_{\odot}$), $P$, the period (in days), and $r_{\rm gd}$, the gas-to-dust ratio, to estimate the progenitor candidate's mass-loss rate assuming it to be in the superwind phase as described above. These authors showed that the prescription does not strongly depend on the gas-to-dust ratio, so we assume a typical value of 200. Then, using the period and luminosity of the progenitor candidate, we obtain its mass-loss rate to be $(2\mbox{--}4)\times10^{-4}~{\rm M}_{\odot}~{\rm yr}^{-1}$, 
which is also within the range obtained from SED modeling by \citet{Jencson_2023ixf}, but lower than the estimate in Paper~II. The latter can be reconciled by adopting a higher value for the wind velocity (e.g., $\sim50$~km/s) appropriate for the superwind phase in the GRAMS SED modeling. The higher velocity does not change the $T_{\rm eff}$ result in Paper~II. Our mass-loss rate estimate here is also in agreement with that obtained by \citet{Grefenstette_2023ixf} based on early hard X-ray observations of SN~2023ixf.

We estimate the initial mass of the progenitor candidate using the Binary Population and Spectral Synthesis (BPASS; \citealt{Stanway-2018}) single-star evolutionary tracks. The metallicity of the host galaxy M101 at the site of the SN~explosion is likely solar (see Paper~II). We therefore use the BPASS stellar models for solar metallicity. Using the $T_{\rm eff}$ range and corresponding luminosity derived above for the RSG progenitor candidate, we plot the latter on the Hertzsprung-Russell (HR) diagram shown in Fig.~\ref{fig:mass}.

We estimate the zero-age main sequence (ZAMS) mass of the progenitor candidate by comparing its luminosity to the terminal points of the tracks. We thus obtain its ZAMS mass of $20\pm4~{\rm M}_{\odot}$. Within the uncertainty, this agrees with the higher end of the mass range obtained from SED fitting with the GRAMS dust models ($17\pm4~{\rm M}_{\odot}$ by \citealt{Jencson_2023ixf}; see also Paper~II). It, however, rules out the lower values  $11\mbox{--}12~{\rm M}_{\odot}$ obtained by \citet{Kilpatrick-2023} and \citet{Pledger-2023}. Our ZAMS mass estimate, therefore, also constrains the different initial-mass results between \citet{Jencson_2023ixf}, Paper~II, and \citet{Kilpatrick-2023}, which are all based on SED fitting, by using an independent method (i.e., the PL relation).

Based on our result above, the progenitor candidate of SN~2023ixf appears to be among the most massive progenitors detected so far for core-collapse SNe. Its inferred ZAMS mass of $20\pm4~{\rm M}_{\odot}$ puts into question the existence of the \emph{red supergiant problem} \citep{Smartt-2015}, i.e., the apparent high-mass cut-off at around $18~{\rm M}_{\odot}$ for RSG stars exploding as SNe despite their high-mass limit of $25\mbox{--}30~{\rm M}_{\odot}$ predicted by stellar evolution theory. 
To-date, a few dozen core-collape SN progenitors have been directly identified (see \citealt{SVD-2017} for a review). Among them, there already exist progenitors with masses close to this limit, for example, SN~2013~ej \citep{SVD-2017}. Furthermore, various other studies, e.g., \citet{Davies-2018, Davies-2020}, have shown that the problem of {\em missing} high-mass progenitors of SNe~II-P/L is not statistically significant and could be explained by the small sample size of the SN progenitors and a steep luminosity function for the RSGs. Our result thus supports this hypothesis.

\subsection{Binary system?}\label{binary}
Other possible origins of variability in luminous near- and mid-IR stellar sources, as described in detail by \citet{Karambelkar_2019}, are the presence of interacting dusty winds in massive (Wolf-Rayet) binary stars and/or orbital modulation of the binary systems. As presented in that work and references therein, such dusty binary systems could be heavily obscured in the  optical, giving a possible explanation for the non-detection of the assumed progenitor system of SN~2023ixf with {\it HST} F435W and F555W filters (see, e.g., \citealt{Soraisam-2023, Kilpatrick-2023}, Paper~II). 
Both the observed IR light-curve amplitudes and period could be explained with the orbital modulation of a massive dusty binary. 
However, based on the results presented above (IR colors, similarity to identified pulsators), a long-period RSG star seems to be the primary candidate for the progenitor of SN~2023ixf. 

On the other hand, we note that \citet{Kilpatrick-2023} also carried out a binary population synthesis analysis using the {\it HST} flux upper limits and found that a close binary as a progenitor system for SN 2023ixf cannot be ruled out. However, although the binary fraction of massive OB-type stars is regarded to be more than 70\% \citep[e.g.,][]{Sana-2012}, recent studies have found a fraction of $20\%$ or less for RSGs with OB-type companions \citep{Neugent-2020}, making this model rather unlikely. As \citet{Kilpatrick-2023} have also noted, the presence of a possible companion star could only be revealed with deep optical imaging after the SN fades.

\section{Conclusions}\label{sec:conclude}
The proximity of SN~2023ixf, along with its host galaxy M101 being a frequent target for various imaging campaigns in different wavelengths, have facilitated detection and detailed studies of its progenitor candidate. Here we used archival mid- ($3.6~\mu{\rm m}$ and $4.5~\mu{\rm m}$) and near-IR $JHK$ data from the {\sl Spitzer Space Telescope} and ground-based facilities (Gemini North, UKIRT, MMT) to characterize the variability of the progenitor candidate, covering a combined duration of around 19~yr prior to the SN explosion. We find its RMS amplitudes in these wavelengths to be in the range of $0.2\mbox{--}0.4$~mag, increasing with decreasing wavelength. From a rigorous period analysis of the higher-cadence {\sl Spitzer} $3.6~\mu{\rm m}$ and $4.5~\mu{\rm m}$ data, we determine a period of $1091\pm71$~d. We demonstrate that the same periodicity is also present in the $JHK$ light curves, thus indicating a common physical origin for the observed variability in both mid- and near-IR. The variability characteristics taken together with the infrared colors are consistent with those of pulsating RSGs.

Using the period-luminosity relation for RSGs from \citet{Soraisam-2018}, we compute an absolute $K$-band magnitude of $-11.58\pm0.31$~mag. 
We then use the range of $T_{\rm eff}$ covered by late M-type RSGs and their corresponding bolometric corrections to derive the luminosity, $\log(L/L_{\odot})$, for the progenitor candidate. 
We obtain a range of $5.27\pm0.12$~dex at $T_{\rm eff} = 3200$~K, the lowest value considered, to $5.37\pm0.12$~dex at $T_{\rm eff} = 3500$~K, the highest value considered.
This thus gives an estimate of the luminosity independent of SED fitting for  the RSG progenitor candidate.  
Assuming the latter to be in the superwind phase (with enhanced dust-driven mass-loss) over the final decades before the explosion \citep[see, e.g.,][]{Jencson_2023ixf} and using its luminosity, pulsation period, and the mass-loss prescription of \citet{Goldman-2017}, we estimate its mass-loss rate. We find a value of $(2\mbox{--}4)\times10^{-4}~{\rm M}_{\odot}~{\rm yr}^{-1}$, in agreement with the results from other authors \citep[e.g.,][]{Grefenstette_2023ixf, Jencson_2023ixf}. Finally, comparing the above luminosity with stellar models, we obtain a zero-age main sequence mass of $20\pm4~M_{\odot}$  for the progenitor candidate of SN~2023ixf, which places it among the most massive progenitors detected so far for type~II~SNe.

\begin{deluxetable*}{cccccc}
\tablewidth{0pt}
\tablecolumns{6}
\tablecaption{{\sl Spitzer\/} IRAC Flux Densities\label{tab:spitzer}}
\tablehead{
\colhead{} & \colhead{} & \multicolumn2c{3.6 $\mu$m} & \multicolumn2c{4.5 $\mu$m} \\
\colhead{} & \colhead{} & \multicolumn2c{\hrulefill} & \multicolumn2c{\hrulefill} \\
\colhead{AORKEY} & \colhead{MJD} & \colhead{Flux Density} & \colhead{Unc.} & \colhead{Flux Density} & \colhead{Unc.} \\
\colhead{} & \colhead{} & \colhead{($\mu$Jy)} & \colhead{($\mu$Jy)} & \colhead{($\mu$Jy)} & \colhead{($\mu$Jy)}
}
\startdata
4370432  &  53072.09  &  36.08  &  4.16  &  32.70  &  4.25 \\
44605952  &  55960.72  &  29.68  &  3.43  &  \nodata  &  \nodata \\
44605696  &  55980.99  &  28.75  &  3.48  &  \nodata  &  \nodata \\
45237760  &  56165.01  &  \nodata  &  \nodata  &  22.27  &  2.52 \\
48186624  &  56337.07  &  31.45  &  3.51  &  \nodata  &  \nodata \\
45237504  &  56348.11  &  30.41  &  3.39  &  \nodata  &  \nodata \\
48187392  &  56516.35  &  \nodata  &  \nodata  &  27.03  &  3.00\\
50627840  &  56742.84  &  43.93  &  4.89  &  37.08  &  4.11 \\
50627328  &  56771.83  &  42.49  &  4.71  &  38.01  &  4.21 \\
50627072  &  56902.01  &  42.06  &  4.75  &  45.96  &  5.08 \\
52776448  &  57136.69  &  26.00  &  2.94  &  23.97  &  2.68 \\
52776704  &  57144.06  &  29.15  &  3.27  &  23.49  &  2.63 \\
52776960  &  57150.17  &  28.40  &  3.20  &  23.75  &  2.66 \\
52777216  &  57163.71  &  27.29  &  3.08  &  23.31  &  2.59 \\
52777472  &  57191.82  &  23.58  &  2.64  &  25.80  &  2.85 \\
52777728  &  57220.79  &  24.33  &  2.75  &  21.91  &  2.44 \\
52777984  &  57247.82  &  25.65  &  2.90  &  23.14  &  2.60 \\
52778240  &  57486.85  &  36.33  &  4.03  &  30.10  &  3.35 \\
60830720  &  57843.93  &  45.05  &  5.00  &  45.28  &  5.01 \\
60830976  &  57926.90  &  \nodata  &  \nodata  &  31.01  &  3.44 \\
60831232  &  58009.67  &  42.32  &  4.70  &  31.07  &  3.44 \\
60831488  &  58232.95  &  31.83  &  3.56  &  22.39  &  2.50 \\
60831744  &  58292.87  &  23.45  &  2.69  &  18.92  &  2.14 \\
60832000  &  58380.22  &  31.04  &  3.47  &  21.24  &  2.37 \\
66022400  &  58572.08  &  36.41  &  4.04  &  35.67  &  3.96 \\
66022656  &  58614.39  &  36.55  &  4.07  &  31.72  &  3.51 \\
66022912  &  58655.68  &  34.52  &  3.85  &  30.88  &  3.43 \\
66023168  &  58697.50  &  47.56  &  5.27  &  33.25  &  3.69 \\
66023424  &  58740.01  &  38.29  &  4.26  &  33.56  &  3.73 \\
66023680  &  58781.31  &  40.92  &  4.55  &  38.02  &  4.21 \\
\enddata
\end{deluxetable*}

\begin{deluxetable*}{cccccccc}
\tablewidth{0pt}
\tablecolumns{8}
\tablecaption{NIR Photometry\label{tab:nir}}
\tablehead{
\colhead{} & \multicolumn2c{$J$} & \multicolumn2c{$H$} & \multicolumn2c{$K$} \\
\colhead{} & \multicolumn2c{\hrulefill} & \multicolumn2c{\hrulefill} & \multicolumn2c{\hrulefill} \\
\colhead{MJD} & \colhead{Mag} & \colhead{Mag\_err} & \colhead{Mag} & \colhead{Mag\_err} & \colhead{Mag} & \colhead{Mag\_err} 
& \colhead{Instrument}
}
\startdata
54230. & 20.39 & 0.24 & 19.18 & 0.09 & 18.25 & 0.05 & WFCAM \\
55304. & 20.12 & 0.10 & \nodata & \nodata & 18.56 & 0.03 & NIRI \\ 
55803. & 20.68 & 0.20 & 19.22 & 0.09 & 18.57 & 0.09 & WFCAM \\
55805. & \nodata & \nodata & \nodata & \nodata & 18.68 & 0.10 & WFCAM \\
55807. & \nodata & \nodata & 19.84 & 0.19 & \nodata & \nodata & WFCAM \\
55899. & \nodata & \nodata & 20.06 & 0.24 & 18.95 & 0.15 & WFCAM \\
55929. & 21.01 & 0.44 & 19.41 & 0.15 & 18.97 & 0.19 & WFCAM \\
56104. & \nodata & \nodata & 20.22 & 0.26 & \nodata & \nodata & WFCAM \\
56399. & 20.95 & 0.19 & 19.46 & 0.05 & 18.88 & 0.09 & WFCAM \\  
\enddata
\tablenotetext{}{All magnitudes in this table are in the Vega system.}
\end{deluxetable*}

\vspace{5mm}
\section*{Acknowledgments}
This work is based in part on archival data obtained with the {\sl Spitzer Space Telescope}, which was operated by the Jet Propulsion Laboratory, California Institute of Technology under a contract with NASA.

Based on observations obtained at the international Gemini Observatory, a program of NSF’s NOIRLab, which is managed by the Association of Universities for Research in Astronomy (AURA) under a cooperative agreement with the National Science Foundation on behalf of the Gemini Observatory partnership: the National Science Foundation (United States), National Research Council (Canada), Agencia Nacional de Investigaci\'{o}n y Desarrollo (Chile), Ministerio de Ciencia, Tecnolog\'{i}a e Innovaci\'{o}n (Argentina), Minist\'{e}rio da Ci\^{e}ncia, Tecnologia, Inova\c{c}\~{o}es e Comunica\c{c}\~{o}es (Brazil), and Korea Astronomy and Space Science Institute (Republic of Korea).

This work was enabled by observations made from the Gemini North  and United Kingdon Infrared (UKIRT) telescopes, located within the Maunakea Science Reserve and adjacent to the summit of Maunakea. We are grateful for the privilege of observing the Universe from a place that is unique in both its astronomical quality and its cultural significance.

UKIRT is owned by the University of Hawaii (UH) and operated by the UH Institute for Astronomy. When the data used here were obtained, UKIRT was operated by the Joint Astronomy Centre on behalf of the Science and Technology Facilities Council of the U.K.

T.S. is supported by the NKFIH/OTKA grant FK-134432 of the National Research, Development and Innovation (NRDI) Office of Hungary, by the János Bolyai Research Scholarship of the Hungarian Academy of Sciences, and by the New National Excellence Program (UNKP-22-5) of the Ministry for Culture and Innovation from the source of the NRDI Fund, Hungary. S.S. and D.A.V.-T. acknowledge support from UNAM-PAPIIT program IA104822. S.H.C. acknowledges support from the National Research Foundation of Korea (NRF) grant funded by the Korea government (MSIT) (NRF-2021R1C1C2003511) and the Korea Astronomy and Space Science Institute under R\&D program (Project No. 2023-1-860-02) supervised by the Ministry of Science and ICT.

\vspace{5mm}
\facilities{{\sl Spitzer}, HST, UKIRT, Gemini North, MMT}

\software{\texttt{numpy} \citep{numpy}, \texttt{scipy} \citep{scipy}, \texttt{astropy} \citep{astropy}, \texttt{matplotlib} \citep{matplotlib}, \texttt{NIFTy} \citep{Selig-2013}}

\clearpage
\appendix
\section{Reconstructed near-IR light curves with modified power spectra}
\label{app}
Here we model the near-IR data of the progenitor candidate using modified versions of the power spectrum recovered from the {\sl Spitzer} $3.6~\mu{\rm m}$ data (Fig.~\ref{fig:spit_recon}, Sect.~\ref{sec:analysis}). The main purpose is to investigate whether we obtain a good reconstruction of the observed data for different shapes of the power spectrum, in particular without the prominent power peak seen in the original spectrum. 

\begin{figure*}
\centering
\includegraphics[width=\textwidth]{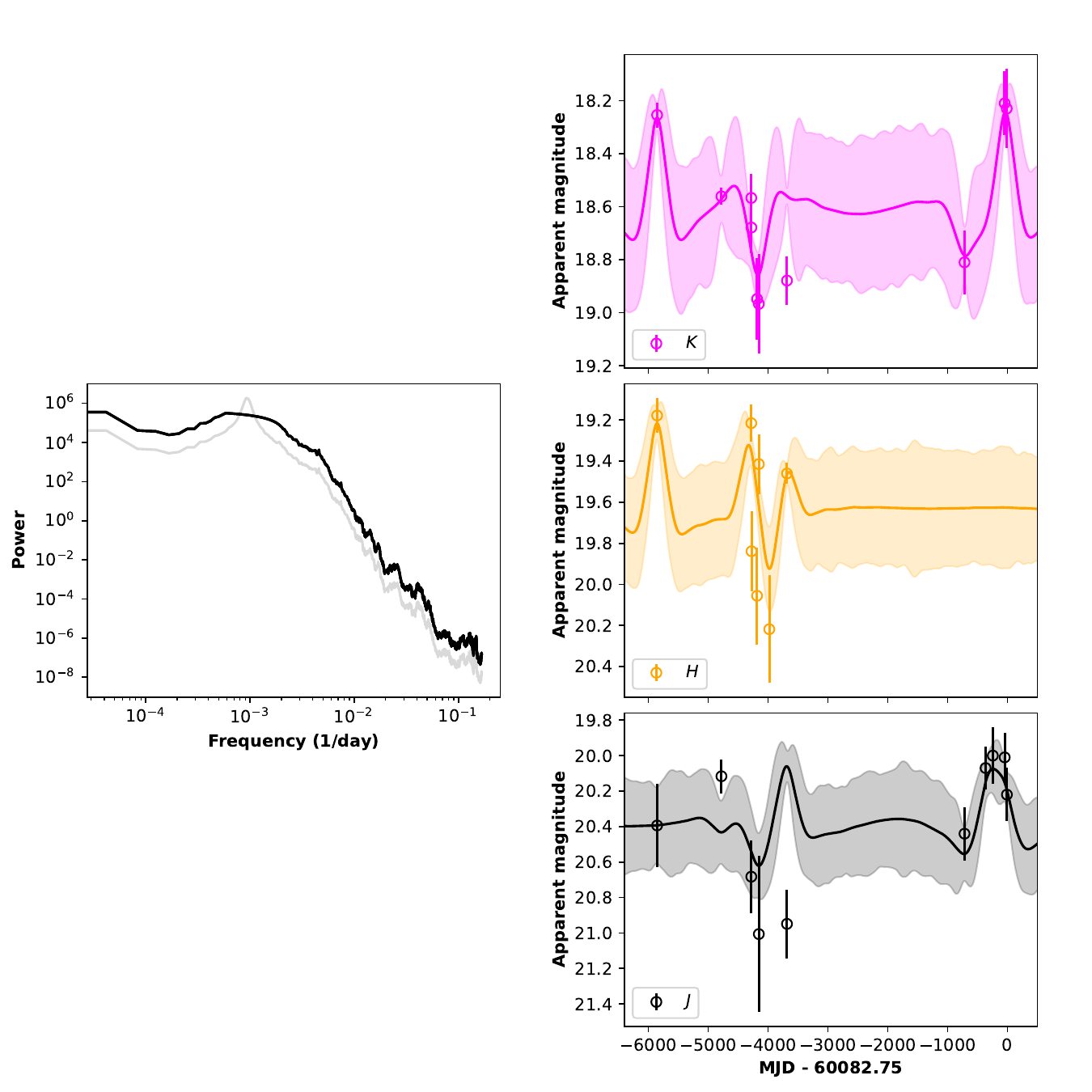}
\caption{Modified form of the power spectrum recovered from the {\sl Spitzer} data ({\it left}, cf.~Sect.~\ref{sec:analysis}) and the near-IR light curve models for the progenitor candidate of SN~2023ixf constructed using it ({\it right}). The grey curve in the left panel shows the original power spectrum, while the shaded regions in the right panel indicate the $1\sigma$ confidence intervals for the light curve models. Here the modified power spectrum is obtained by interpolating across the power peak and scaling it to conserve the total power.}
\label{fig:bad_2}
\end{figure*}

\begin{figure*}
\centering
\includegraphics[width=\textwidth]{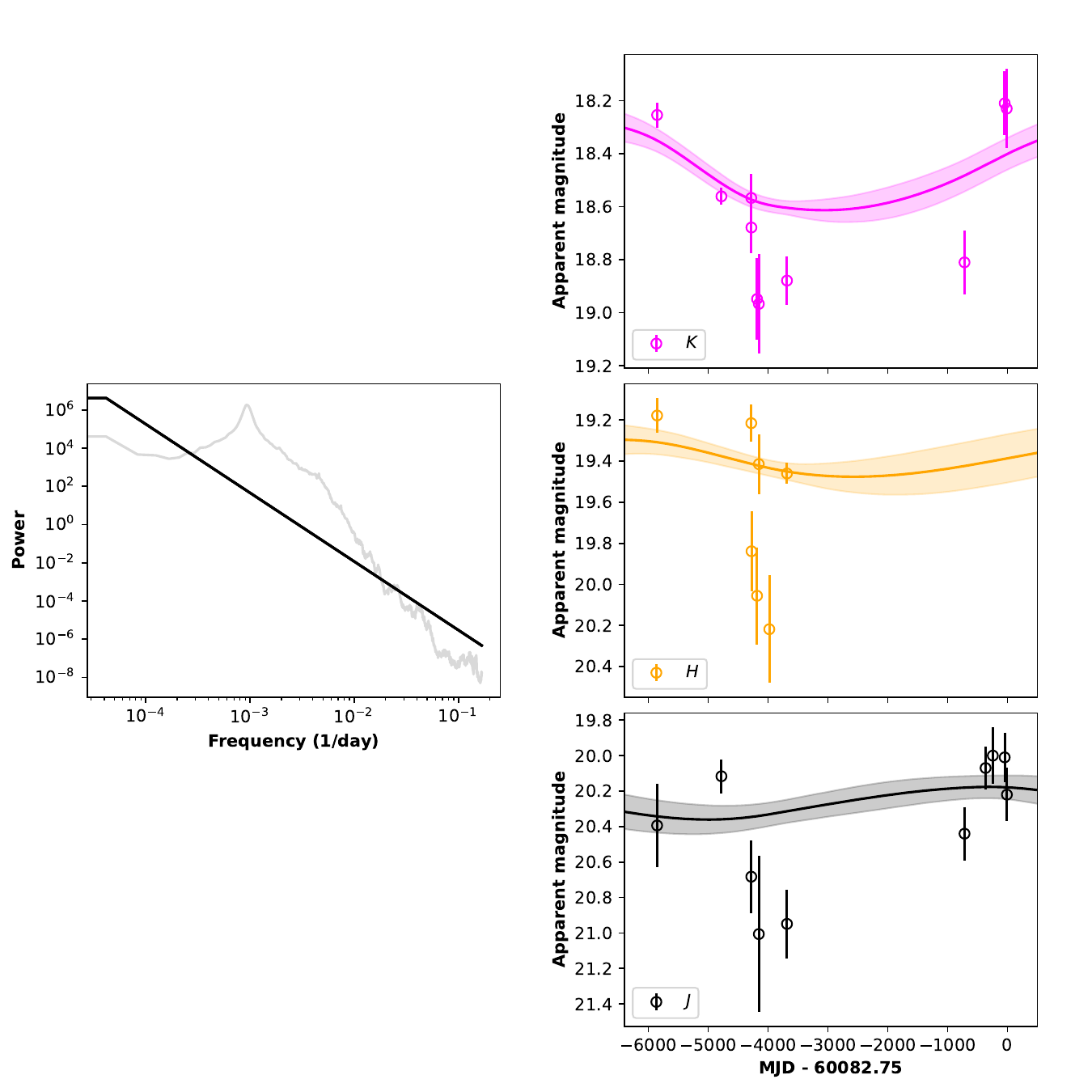}
\caption{Similar to Fig.~\ref{fig:bad_2}. Here the modified power spectrum is obtained by fitting a power law to the original spectrum from the {\sl Spitzer} data and scaling it to conserve the total power. The flattening at the lowest frequency is due to extrapolation using the nearest neighbor value.}
\label{fig:bad_3}
\end{figure*}

\begin{figure*}
\centering
\includegraphics[width=\textwidth]{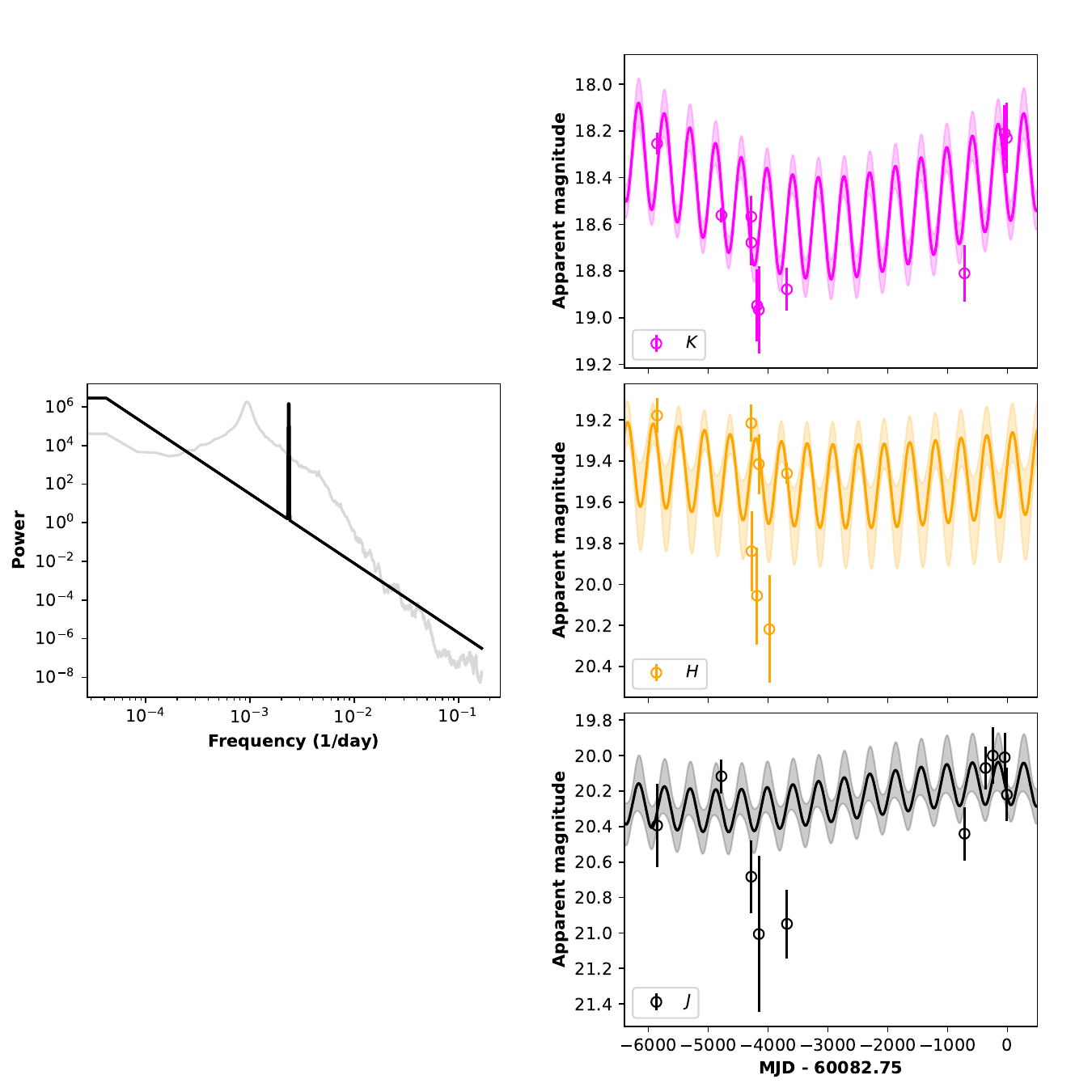}
\caption{Similar to Fig.~\ref{fig:bad_2}. Here the modified power spectrum is obtained by fitting a power law to the original spectrum from the {\sl Spitzer} data, adding a peak at some higher frequency (which is not a harmonic of the original peak frequency) and scaling the resulting spectrum to conserve the total power. The flattening at the lowest frequency is due to extrapolation using the nearest neighbor value.}
\label{fig:bad_4}
\end{figure*}

We consider three flavors -- (i) the same spectrum from Fig.~\ref{fig:nir_recon} but interpolating across the power peak to remove it (Fig.~\ref{fig:bad_2}), (ii) a power law fit to the original spectrum (Fig.~\ref{fig:bad_3}), and (iii) a power peak superposed on this power law fit at some higher frequency that is not a harmonic of the original peak frequency (Fig.~\ref{fig:bad_4}). As can be seen, in all three cases we obtain a poorer reconstruction of the near-IR light curves than with the original power spectrum (Fig.~\ref{fig:nir_recon}), thus demonstrating that modeling the near-IR data indeed requires excess power at the same frequencies as inferred for the mid-IR.

\bibliography{sn23ixf_progenitor_variable}{}

\begin{thebibliography}{}
\expandafter\ifx\csname natexlab\endcsname\relax\def\natexlab#1{#1}\fi
\providecommand{\url}[1]{\href{#1}{#1}}
\providecommand{\dodoi}[1]{doi:~\href{http://doi.org/#1}{\nolinkurl{#1}}}
\providecommand{\doeprint}[1]{\href{http://ascl.net/#1}{\nolinkurl{http://ascl.net/#1}}}
\providecommand{\doarXiv}[1]{\href{https://arxiv.org/abs/#1}{\nolinkurl{https://arxiv.org/abs/#1}}}

\bibitem[{{Arras} {et~al.}(2019){Arras}, {Baltac}, {Ensslin}, {Frank},
  {Hutschenreuter}, {Knollmueller}, {Leike}, {Newrzella}, {Platz}, {Reinecke},
  \& {Stadler}}]{nifty-5}
{Arras}, P., {Baltac}, M., {Ensslin}, T.~A., {et~al.} 2019, {NIFTy5: Numerical
  Information Field Theory v5}, Astrophysics Source Code Library, record
  ascl:1903.008.
\newblock \doeprint{1903.008}

\bibitem[{{Astropy Collaboration} {et~al.}(2013){Astropy Collaboration},
  {Robitaille}, {Tollerud}, {Greenfield}, {Droettboom}, {Bray}, {Aldcroft},
  {Davis}, {Ginsburg}, {Price-Whelan}, {Kerzendorf}, {Conley}, {Crighton},
  {Barbary}, {Muna}, {Ferguson}, {Grollier}, {Parikh}, {Nair}, {Unther},
  {Deil}, {Woillez}, {Conseil}, {Kramer}, {Turner}, {Singer}, {Fox}, {Weaver},
  {Zabalza}, {Edwards}, {Azalee Bostroem}, {Burke}, {Casey}, {Crawford},
  {Dencheva}, {Ely}, {Jenness}, {Labrie}, {Lim}, {Pierfederici}, {Pontzen},
  {Ptak}, {Refsdal}, {Servillat}, \& {Streicher}}]{astropy}
{Astropy Collaboration}, {Robitaille}, T.~P., {Tollerud}, E.~J., {et~al.} 2013,
  \aap, 558, A33, \dodoi{10.1051/0004-6361/201322068}

\bibitem[{{Bellm} {et~al.}(2019){Bellm}, {Kulkarni}, {Graham}, {Dekany},
  {Smith}, {Riddle}, {Masci}, {Helou}, {Prince}, {Adams}, {Barbarino},
  {Barlow}, {Bauer}, {Beck}, {Belicki}, {Biswas}, {Blagorodnova}, {Bodewits},
  {Bolin}, {Brinnel}, {Brooke}, {Bue}, {Bulla}, {Burruss}, {Cenko}, {Chang},
  {Connolly}, {Coughlin}, {Cromer}, {Cunningham}, {De}, {Delacroix}, {Desai},
  {Duev}, {Eadie}, {Farnham}, {Feeney}, {Feindt}, {Flynn}, {Franckowiak},
  {Frederick}, {Fremling}, {Gal-Yam}, {Gezari}, {Giomi}, {Goldstein},
  {Golkhou}, {Goobar}, {Groom}, {Hacopians}, {Hale}, {Henning}, {Ho}, {Hover},
  {Howell}, {Hung}, {Huppenkothen}, {Imel}, {Ip}, {Ivezi{\'c}}, {Jackson},
  {Jones}, {Juric}, {Kasliwal}, {Kaspi}, {Kaye}, {Kelley}, {Kowalski},
  {Kramer}, {Kupfer}, {Landry}, {Laher}, {Lee}, {Lin}, {Lin}, {Lunnan},
  {Giomi}, {Mahabal}, {Mao}, {Miller}, {Monkewitz}, {Murphy}, {Ngeow},
  {Nordin}, {Nugent}, {Ofek}, {Patterson}, {Penprase}, {Porter}, {Rauch},
  {Rebbapragada}, {Reiley}, {Rigault}, {Rodriguez}, {van Roestel}, {Rusholme},
  {van Santen}, {Schulze}, {Shupe}, {Singer}, {Soumagnac}, {Stein}, {Surace},
  {Sollerman}, {Szkody}, {Taddia}, {Terek}, {Van Sistine}, {van Velzen},
  {Vestrand}, {Walters}, {Ward}, {Ye}, {Yu}, {Yan}, \& {Zolkower}}]{Bellm-2019}
{Bellm}, E.~C., {Kulkarni}, S.~R., {Graham}, M.~J., {et~al.} 2019, \pasp, 131,
  018002, \dodoi{10.1088/1538-3873/aaecbe}

\bibitem[{{Berger} {et~al.}(2023){Berger}, {Keating}, {Alexander}, {Cendes},
  {Eftekhari}, {Gurwell}, {Hiramatsu}, {Ho}, {Laskar}, {Margutti}, {Rao}, \&
  {Williams}}]{Berger_2023ixf}
{Berger}, E., {Keating}, G., {Alexander}, K., {et~al.} 2023, Transient Name
  Server AstroNote, 131, 1

\bibitem[{{Bond} {et~al.}(2022){Bond}, {Jencson}, {Whitelock}, {Adams},
  {Bally}, {Cody}, {Gehrz}, {Kasliwal}, \& {Masci}}]{Bond_2022}
{Bond}, H.~E., {Jencson}, J.~E., {Whitelock}, P.~A., {et~al.} 2022, \apj, 928,
  158, \dodoi{10.3847/1538-4357/ac5832}

\bibitem[{{Boyer} {et~al.}(2015){Boyer}, {McQuinn}, {Barmby}, {Bonanos},
  {Gehrz}, {Gordon}, {Groenewegen}, {Lagadec}, {Lennon}, {Marengo}, {McDonald},
  {Meixner}, {Skillman}, {Sloan}, {Sonneborn}, {van Loon}, \&
  {Zijlstra}}]{Boyer_2015}
{Boyer}, M.~L., {McQuinn}, K. B.~W., {Barmby}, P., {et~al.} 2015, \apj, 800,
  51, \dodoi{10.1088/0004-637X/800/1/51}

\bibitem[{Bradley {et~al.}(2022)Bradley, Sipőcz, Robitaille, Tollerud,
  Vinícius, Deil, Barbary, Wilson, Busko, Donath, Günther, Cara, Lim,
  Meßlinger, Conseil, Bostroem, Droettboom, Bray, Bratholm, Barentsen, Craig,
  Rathi, Pascual, Perren, Georgiev, de~Val-Borro, Kerzendorf, Bach, Quint, \&
  Souchereau}]{photutils}
Bradley, L., Sipőcz, B., Robitaille, T., {et~al.} 2022, astropy/photutils:
  1.5.0, 1.5.0,  Zenodo, \dodoi{10.5281/zenodo.6825092}

\bibitem[{{Carey} {et~al.}(2012){Carey}, {Ingalls}, {Hora}, {Surace},
  {Glaccum}, {Lowrance}, {Krick}, {Cole}, {Laine}, {Engelke}, {Price},
  {Bohlin}, \& {Gordon}}]{Carey2012}
{Carey}, S., {Ingalls}, J., {Hora}, J., {et~al.} 2012, in Society of
  Photo-Optical Instrumentation Engineers (SPIE) Conference Series, Vol. 8442,
  Space Telescopes and Instrumentation 2012: Optical, Infrared, and Millimeter
  Wave, ed. M.~C. {Clampin}, G.~G. {Fazio}, H.~A. {MacEwen}, \& J.~{Oschmann},
  Jacobus~M., 84421Z, \dodoi{10.1117/12.927183}

\bibitem[{{Casali} {et~al.}(2007){Casali}, {Adamson}, {Alves de Oliveira},
  {Almaini}, {Burch}, {Chuter}, {Elliot}, {Folger}, {Foucaud}, {Hambly},
  {Hastie}, {Henry}, {Hirst}, {Irwin}, {Ives}, {Lawrence}, {Laidlaw}, {Lee},
  {Lewis}, {Lunney}, {McLay}, {Montgomery}, {Pickup}, {Read}, {Rees}, {Robson},
  {Sekiguchi}, {Vick}, {Warren}, \& {Woodward}}]{wfcam}
{Casali}, M., {Adamson}, A., {Alves de Oliveira}, C., {et~al.} 2007, \aap, 467,
  777, \dodoi{10.1051/0004-6361:20066514}

\bibitem[{{Chandra} {et~al.}(2023){Chandra}, {Maeda}, {Chevalier}, {Nayana}, \&
  {Ray}}]{Chandra_2023ixf_ATel}
{Chandra}, P., {Maeda}, K., {Chevalier}, R.~A., {Nayana}, A.~J., \& {Ray}, A.
  2023, The Astronomer's Telegram, 16073, 1

\bibitem[{{Chatys} {et~al.}(2019){Chatys}, {Bedding}, {Murphy}, {Kiss},
  {Dobie}, \& {Grindlay}}]{Chatys-2019}
{Chatys}, F.~W., {Bedding}, T.~R., {Murphy}, S.~J., {et~al.} 2019, \mnras, 487,
  4832, \dodoi{10.1093/mnras/stz1584}

\bibitem[{{Davies} \& {Beasor}(2018)}]{Davies-2018}
{Davies}, B., \& {Beasor}, E.~R. 2018, \mnras, 474, 2116,
  \dodoi{10.1093/mnras/stx2734}

\bibitem[{{Davies} \& {Beasor}(2020)}]{Davies-2020}
---. 2020, \mnras, 493, 468, \dodoi{10.1093/mnras/staa174}

\bibitem[{{Davies} {et~al.}(2022){Davies}, {Plez}, \& {Petrault}}]{Davies-2022}
{Davies}, B., {Plez}, B., \& {Petrault}, M. 2022, \mnras, 517, 1483,
  \dodoi{10.1093/mnras/stac2427}

\bibitem[{{Dong} {et~al.}(2023){Dong}, {Sand}, {Valenti}, {Bostroem},
  {Andrews}, {Hosseinzadeh}, {Hoang}, {Janzen}, {Jencson}, {Lundquist}, {Meza
  Retamal}, {Pearson}, {Shrestha}, {Haislip}, {Kouprianov}, \&
  {Reichart}}]{Dong-2023}
{Dong}, Y., {Sand}, D.~J., {Valenti}, S., {et~al.} 2023, arXiv e-prints,
  arXiv:2307.02539, \dodoi{10.48550/arXiv.2307.02539}

\bibitem[{{Goldman} {et~al.}(2017){Goldman}, {van Loon}, {Zijlstra}, {Green},
  {Wood}, {Nanni}, {Imai}, {Whitelock}, {Matsuura}, {Groenewegen}, \&
  {G{\'o}mez}}]{Goldman-2017}
{Goldman}, S.~R., {van Loon}, J.~T., {Zijlstra}, A.~A., {et~al.} 2017, \mnras,
  465, 403, \dodoi{10.1093/mnras/stw2708}

\bibitem[{{Grefenstette} {et~al.}(2023){Grefenstette}, {Brightman}, {Earnshaw},
  {Harrison}, \& {Margutti}}]{Grefenstette_2023ixf}
{Grefenstette}, B.~W., {Brightman}, M., {Earnshaw}, H.~P., {Harrison}, F.~A.,
  \& {Margutti}, R. 2023, arXiv e-prints, arXiv:2306.04827,
  \dodoi{10.48550/arXiv.2306.04827}

\bibitem[{{Guo} \& {Li}(2002)}]{Guo-2002}
{Guo}, J.~H., \& {Li}, Y. 2002, \apj, 565, 559, \dodoi{10.1086/324295}

\bibitem[{{Gvaramadze} {et~al.}(2014){Gvaramadze}, {Menten}, {Kniazev},
  {Langer}, {Mackey}, {Kraus}, {Meyer}, \& {Kami{\'n}ski}}]{Gvaramadze-2014}
{Gvaramadze}, V.~V., {Menten}, K.~M., {Kniazev}, A.~Y., {et~al.} 2014, \mnras,
  437, 843, \dodoi{10.1093/mnras/stt1943}

\bibitem[{{Hambly} {et~al.}(2008){Hambly}, {Collins}, {Cross}, {Mann}, {Read},
  {Sutorius}, {Bond}, {Bryant}, {Emerson}, {Lawrence}, {Rimoldini}, {Stewart},
  {Williams}, {Adamson}, {Hirst}, {Dye}, \& {Warren}}]{ukirt_archive}
{Hambly}, N.~C., {Collins}, R.~S., {Cross}, N.~J.~G., {et~al.} 2008, \mnras,
  384, 637, \dodoi{10.1111/j.1365-2966.2007.12700.x}

\bibitem[{Harris {et~al.}(2020)Harris, Millman, van~der Walt, Gommers,
  Virtanen, Cournapeau, Wieser, Taylor, Berg, Smith, Kern, Picus, Hoyer, van
  Kerkwijk, Brett, Haldane, del R{\'{i}}o, Wiebe, Peterson,
  G{\'{e}}rard-Marchant, Sheppard, Reddy, Weckesser, Abbasi, Gohlke, \&
  Oliphant}]{numpy}
Harris, C.~R., Millman, K.~J., van~der Walt, S.~J., {et~al.} 2020, Nature, 585,
  357, \dodoi{10.1038/s41586-020-2649-2}

\bibitem[{{Heger} {et~al.}(1997){Heger}, {Jeannin}, {Langer}, \&
  {Baraffe}}]{Heger-1997}
{Heger}, A., {Jeannin}, L., {Langer}, N., \& {Baraffe}, I. 1997, \aap, 327,
  224, \dodoi{10.48550/arXiv.astro-ph/9705097}

\bibitem[{{Hirst} \& {Cardenes}(2016)}]{gem_archive}
{Hirst}, P., \& {Cardenes}, R. 2016, in Society of Photo-Optical
  Instrumentation Engineers (SPIE) Conference Series, Vol. 9913, Software and
  Cyberinfrastructure for Astronomy IV, ed. G.~{Chiozzi} \& J.~C. {Guzman},
  99131E, \dodoi{10.1117/12.2231833}

\bibitem[{{Hodapp} {et~al.}(2003){Hodapp}, {Jensen}, {Irwin}, {Yamada},
  {Chung}, {Fletcher}, {Robertson}, {Hora}, {Simons}, {Mays}, {Nolan}, {Bec},
  {Merrill}, \& {Fowler}}]{niri}
{Hodapp}, K.~W., {Jensen}, J.~B., {Irwin}, E.~M., {et~al.} 2003, \pasp, 115,
  1388, \dodoi{10.1086/379669}

\bibitem[{{Hodgkin} {et~al.}(2009){Hodgkin}, {Irwin}, {Hewett}, \&
  {Warren}}]{Hodgkin-2009}
{Hodgkin}, S.~T., {Irwin}, M.~J., {Hewett}, P.~C., \& {Warren}, S.~J. 2009,
  \mnras, 394, 675, \dodoi{10.1111/j.1365-2966.2008.14387.x}

\bibitem[{{Hosseinzadeh} {et~al.}(2023){Hosseinzadeh}, {Farah}, {Shrestha},
  {Sand}, {Dong}, {Brown}, {Bostroem}, {Valenti}, {Jha}, {Andrews}, {Arcavi},
  {Haislip}, {Hiramatsu}, {Hoang}, {Howell}, {Janzen}, {Jencson}, {Kouprianov},
  {Lundquist}, {McCully}, {Meza Retamal}, {Modjaz}, {Newsome}, {Padilla
  Gonzalez}, {Pearson}, {Pellegrino}, {Ravi}, {Reichart}, {Smith}, {Terreran},
  \& {Vink{\'o}}}]{Hosseinzadeh_2023ixf}
{Hosseinzadeh}, G., {Farah}, J., {Shrestha}, M., {et~al.} 2023, arXiv e-prints,
  arXiv:2306.06097, \dodoi{10.48550/arXiv.2306.06097}

\bibitem[{Hunter(2007)}]{matplotlib}
Hunter, J.~D. 2007, Computing In Science \& Engineering, 9, 90,
  \dodoi{10.1109/MCSE.2007.55}

\bibitem[{{Iben} \& {Renzini}(1983)}]{Iben-1983}
{Iben}, I., J., \& {Renzini}, A. 1983, \araa, 21, 271,
  \dodoi{10.1146/annurev.aa.21.090183.001415}

\bibitem[{{Irwin} {et~al.}(2004){Irwin}, {Lewis}, {Hodgkin}, {Bunclark},
  {Evans}, {McMahon}, {Emerson}, {Stewart}, \& {Beard}}]{ukirt_pipeline}
{Irwin}, M.~J., {Lewis}, J., {Hodgkin}, S., {et~al.} 2004, in Society of
  Photo-Optical Instrumentation Engineers (SPIE) Conference Series, Vol. 5493,
  Optimizing Scientific Return for Astronomy through Information Technologies,
  ed. P.~J. {Quinn} \& A.~{Bridger}, 411--422, \dodoi{10.1117/12.551449}

\bibitem[{{Itagaki}(2023)}]{Itagaki}
{Itagaki}, K. 2023, Transient Name Server Discovery Report, 2023-1158, 1

\bibitem[{{Jacobson-Galan} {et~al.}(2023){Jacobson-Galan}, {Dessart},
  {Margutti}, {Chornock}, {Foley}, {Kilpatrick}, {Jones}, {Taggart}, {Angus},
  {Bhattacharjee}, {Braff}, {Brethauer}, {Burgasser}, {Cao}, {Carlile},
  {Chambers}, {Coulter}, {Dominguez-Ruiz}, {Dickinson}, {de Boer}, {Gagliano},
  {Gall}, {Gao}, {Gates}, {Gomez}, {Guolo}, {Halford}, {Hjorth}, {Huber},
  {Johnson}, {Karpoor}, {Laskar}, {LeBaron}, {Li}, {Lin}, {Loch}, {Lynam},
  {Magnier}, {Maloney}, {Matthews}, {McDonald}, {Miao}, {Milisavljevic}, {Pan},
  {Pradyumna}, {Ransome}, {Rees}, {Rest}, {Rojas-Bravo}, {Sandford}, {Sandoval
  Ascencio}, {Sanjaripour}, {Savino}, {Sears}, {Sharei}, {Smartt}, {Softich},
  {Theissen}, {Tinyanont}, {Tohfa}, {Villar}, {Wang}, {Wainscoat},
  {Westerling}, {Wiston}, {Wozniak}, {Yadavalli}, \&
  {Zenati}}]{Jacobson-Galan_2023ixf}
{Jacobson-Galan}, W.~V., {Dessart}, L., {Margutti}, R., {et~al.} 2023, arXiv
  e-prints, arXiv:2306.04721, \dodoi{10.48550/arXiv.2306.04721}

\bibitem[{{Jencson} {et~al.}(2023){Jencson}, {Pearson}, {Beasor}, {Lau},
  {Andrews}, {Bostroem}, {Dong}, {Engesser}, {Gomez}, {Guolo}, {Hoang},
  {Hosseinzadeh}, {Jha}, {Karambelkar}, {Kasliwal}, {Lundquist}, {Meza
  Retamal}, {Rest}, {Sand}, {Shahbandeh}, {Shrestha}, {Smith}, {Strader},
  {Valenti}, {Wang}, \& {Zenati}}]{Jencson_2023ixf}
{Jencson}, J.~E., {Pearson}, J., {Beasor}, E.~R., {et~al.} 2023, arXiv
  e-prints, arXiv:2306.08678.
\newblock \doarXiv{2306.08678}

\bibitem[{{Jurcevic} {et~al.}(2000){Jurcevic}, {Pierce}, \&
  {Jacoby}}]{Jurcevic-2000}
{Jurcevic}, J.~S., {Pierce}, M.~J., \& {Jacoby}, G.~H. 2000, \mnras, 313, 868,
  \dodoi{10.1046/j.1365-8711.2000.03292.x}

\bibitem[{{Justtanont} {et~al.}(2013){Justtanont}, {Teyssier}, {Barlow},
  {Matsuura}, {Swinyard}, {Waters}, \& {Yates}}]{Justtanont-2013}
{Justtanont}, K., {Teyssier}, D., {Barlow}, M.~J., {et~al.} 2013, \aap, 556,
  A101, \dodoi{10.1051/0004-6361/201321812}

\bibitem[{{Karambelkar} {et~al.}(2019){Karambelkar}, {Adams}, {Whitelock},
  {Kasliwal}, {Jencson}, {Boyer}, {Goldman}, {Masci}, {Cody}, {Bally}, {Bond},
  {Gehrz}, {Parthasarathy}, {Lau}, \& {SPIRITS
  Collaboration}}]{Karambelkar_2019}
{Karambelkar}, V.~R., {Adams}, S.~M., {Whitelock}, P.~A., {et~al.} 2019, \apj,
  877, 110, \dodoi{10.3847/1538-4357/ab1a41}

\bibitem[{{Kasliwal} {et~al.}(2017){Kasliwal}, {Bally}, {Masci}, {Cody},
  {Bond}, {Jencson}, {Tinyanont}, {Cao}, {Contreras}, {Dykhoff}, {Amodeo},
  {Armus}, {Boyer}, {Cantiello}, {Carlon}, {Cass}, {Cook}, {Corgan}, {Faella},
  {Fox}, {Green}, {Gehrz}, {Helou}, {Hsiao}, {Johansson}, {Khan}, {Lau},
  {Langer}, {Levesque}, {Milne}, {Mohamed}, {Morrell}, {Monson}, {Moore},
  {Ofek}, {O' Sullivan}, {Parthasarathy}, {Perez}, {Perley}, {Phillips},
  {Prince}, {Shenoy}, {Smith}, {Surace}, {Van Dyk}, {Whitelock}, \&
  {Williams}}]{Kasliwal2017}
{Kasliwal}, M.~M., {Bally}, J., {Masci}, F., {et~al.} 2017, \apj, 839, 88,
  \dodoi{10.3847/1538-4357/aa6978}

\bibitem[{{Kilpatrick} {et~al.}(2023){Kilpatrick}, {Foley},
  {Jacobson-Gal{\'a}n}, {Piro}, {Smartt}, {Drout}, {Gagliano}, {Gall},
  {Hjorth}, {Jones}, {Mandel}, {Margutti}, {Ransome}, {Villar}, {Coulter},
  {Gao}, {Matthews}, \& {Zenati}}]{Kilpatrick-2023}
{Kilpatrick}, C.~D., {Foley}, R.~J., {Jacobson-Gal{\'a}n}, W.~V., {et~al.}
  2023, arXiv e-prints, arXiv:2306.04722, \dodoi{10.48550/arXiv.2306.04722}

\bibitem[{{Kiss} {et~al.}(2006){Kiss}, {Szab{\'o}}, \& {Bedding}}]{Kiss-2006}
{Kiss}, L.~L., {Szab{\'o}}, G.~M., \& {Bedding}, T.~R. 2006, \mnras, 372, 1721,
  \dodoi{10.1111/j.1365-2966.2006.10973.x}

\bibitem[{{Labrie} {et~al.}(2019){Labrie}, {Anderson}, {C{\'a}rdenes},
  {Simpson}, \& {Turner}}]{dragons}
{Labrie}, K., {Anderson}, K., {C{\'a}rdenes}, R., {Simpson}, C., \& {Turner},
  J. E.~H. 2019, in Astronomical Society of the Pacific Conference Series, Vol.
  523, Astronomical Data Analysis Software and Systems XXVII, ed. P.~J.
  {Teuben}, M.~W. {Pound}, B.~A. {Thomas}, \& E.~M. {Warner}, 321

\bibitem[{{Levesque} {et~al.}(2005){Levesque}, {Massey}, {Olsen}, {Plez},
  {Josselin}, {Maeder}, \& {Meynet}}]{Levesque-2005}
{Levesque}, E.~M., {Massey}, P., {Olsen}, K.~A.~G., {et~al.} 2005, \apj, 628,
  973, \dodoi{10.1086/430901}

\bibitem[{{Levesque} {et~al.}(2009){Levesque}, {Massey}, {Plez}, \&
  {Olsen}}]{Levesque-2009}
{Levesque}, E.~M., {Massey}, P., {Plez}, B., \& {Olsen}, K. A.~G. 2009, \aj,
  137, 4744, \dodoi{10.1088/0004-6256/137/6/4744}

\bibitem[{{Makovoz} \& {Khan}(2005)}]{Makovoz2005a}
{Makovoz}, D., \& {Khan}, I. 2005, in Astronomical Society of the Pacific
  Conference Series, Vol. 347, Astronomical Data Analysis Software and Systems
  XIV, ed. P.~{Shopbell}, M.~{Britton}, \& R.~{Ebert}, 81

\bibitem[{{Masci} {et~al.}(2019){Masci}, {Laher}, {Rusholme}, {Shupe}, {Groom},
  {Surace}, {Jackson}, {Monkewitz}, {Beck}, {Flynn}, {Terek}, {Landry},
  {Hacopians}, {Desai}, {Howell}, {Brooke}, {Imel}, {Wachter}, {Ye}, {Lin},
  {Cenko}, {Cunningham}, {Rebbapragada}, {Bue}, {Miller}, {Mahabal}, {Bellm},
  {Patterson}, {Juri{\'c}}, {Golkhou}, {Ofek}, {Walters}, {Graham}, {Kasliwal},
  {Dekany}, {Kupfer}, {Burdge}, {Cannella}, {Barlow}, {Van Sistine}, {Giomi},
  {Fremling}, {Blagorodnova}, {Levitan}, {Riddle}, {Smith}, {Helou}, {Prince},
  \& {Kulkarni}}]{Masci-2019}
{Masci}, F.~J., {Laher}, R.~R., {Rusholme}, B., {et~al.} 2019, \pasp, 131,
  018003, \dodoi{10.1088/1538-3873/aae8ac}

\bibitem[{{Massey} {et~al.}(2023){Massey}, {Neugent}, {Ekstr{\"o}m}, {Georgy},
  \& {Meynet}}]{Massey_2023}
{Massey}, P., {Neugent}, K.~F., {Ekstr{\"o}m}, S., {Georgy}, C., \& {Meynet},
  G. 2023, \apj, 942, 69, \dodoi{10.3847/1538-4357/aca665}

\bibitem[{{Matheson} {et~al.}(2021){Matheson}, {Stubens}, {Wolf}, {Lee},
  {Narayan}, {Saha}, {Scott}, {Soraisam}, {Bolton}, {Hauger}, {Silva},
  {Kececioglu}, {Scheidegger}, {Snodgrass}, {Aleo}, {Evans-Jacquez}, {Singh},
  {Wang}, {Yang}, \& {Zhao}}]{Antares}
{Matheson}, T., {Stubens}, C., {Wolf}, N., {et~al.} 2021, \aj, 161, 107,
  \dodoi{10.3847/1538-3881/abd703}

\bibitem[{{Matthews} {et~al.}(2023{\natexlab{a}}){Matthews}, {Margutti},
  {Alexander}, {Bright}, {Cendes}, {Berger}, {Lasker}, {Drout}, \&
  {Milisavljevic}}]{Matthews_2023ixf_TNS}
{Matthews}, D., {Margutti}, R., {Alexander}, K.~D., {et~al.}
  2023{\natexlab{a}}, Transient Name Server AstroNote, 146, 1

\bibitem[{{Matthews} {et~al.}(2023{\natexlab{b}}){Matthews}, {Margutti}, {AJ},
  {Jacobson-Galan}, {Chornock}, {Alexander}, {Laskar}, {Cendes}, {Berger},
  {Drout}, \& {Milisavljevic}}]{Matthews-2023}
{Matthews}, D., {Margutti}, R., {AJ}, N., {et~al.} 2023{\natexlab{b}}, The
  Astronomer's Telegram, 16091, 1

\bibitem[{{Mauron} \& {Josselin}(2011)}]{Mauron-2011}
{Mauron}, N., \& {Josselin}, E. 2011, \aap, 526, A156,
  \dodoi{10.1051/0004-6361/201013993}

\bibitem[{{Mereminskiy} {et~al.}(2023){Mereminskiy}, {Lutovinov}, {Sazonov},
  {Arefiev}, {Lapshov}, {Molkov}, {Semena}, {Shtykovsky}, \&
  {Tkachenko}}]{Mereminskiy_2023ixf_ATel}
{Mereminskiy}, I.~A., {Lutovinov}, A.~A., {Sazonov}, S.~Y., {et~al.} 2023, The
  Astronomer's Telegram, 16065, 1

\bibitem[{{Neugent} {et~al.}(2020){Neugent}, {Levesque}, {Massey}, {Morrell},
  \& {Drout}}]{Neugent-2020}
{Neugent}, K.~F., {Levesque}, E.~M., {Massey}, P., {Morrell}, N.~I., \&
  {Drout}, M.~R. 2020, \apj, 900, 118, \dodoi{10.3847/1538-4357/ababaa}

\bibitem[{{Neustadt} {et~al.}(2023){Neustadt}, {Kochanek}, \& {Rizzo
  Smith}}]{Neustadt-2023}
{Neustadt}, J.~M.~M., {Kochanek}, C.~S., \& {Rizzo Smith}, M. 2023, arXiv
  e-prints, arXiv:2306.06162, \dodoi{10.48550/arXiv.2306.06162}

\bibitem[{{O'Grady} {et~al.}(2020){O'Grady}, {Drout}, {Shappee}, {Bauer},
  {Fuller}, {Kochanek}, {Jayasinghe}, {Gaensler}, {Stanek}, {Holoien},
  {Prieto}, \& {Thompson}}]{Grady-2020}
{O'Grady}, A. J.~G., {Drout}, M.~R., {Shappee}, B.~J., {et~al.} 2020, \apj,
  901, 135, \dodoi{10.3847/1538-4357/abafad}

\bibitem[{{Perley} {et~al.}(2023){Perley}, {Gal-Yam}, {Irani}, \&
  {Zimmerman}}]{Perley_2023ixf}
{Perley}, D.~A., {Gal-Yam}, A., {Irani}, I., \& {Zimmerman}, E. 2023, Transient
  Name Server AstroNote, 119, 1

\bibitem[{{Perley} \& {Irani}(2023)}]{ZTF_2023ixf}
{Perley}, D.~A., \& {Irani}, I. 2023, Transient Name Server AstroNote, 120, 1

\bibitem[{{Pledger} \& {Shara}(2023)}]{Pledger-2023}
{Pledger}, J.~L., \& {Shara}, M.~M. 2023, arXiv e-prints, arXiv:2305.14447,
  \dodoi{10.48550/arXiv.2305.14447}

\bibitem[{{Reiter} {et~al.}(2015){Reiter}, {Marengo}, {Hora}, \&
  {Fazio}}]{Reiter_2015}
{Reiter}, M., {Marengo}, M., {Hora}, J.~L., \& {Fazio}, G.~G. 2015, \mnras,
  447, 3909, \dodoi{10.1093/mnras/stu2725}

\bibitem[{{Ren} {et~al.}(2019){Ren}, {Jiang}, {Yang}, \& {Gao}}]{Ren-2019}
{Ren}, Y., {Jiang}, B.-W., {Yang}, M., \& {Gao}, J. 2019, \apjs, 241, 35,
  \dodoi{10.3847/1538-4365/ab0825}

\bibitem[{{Sana} {et~al.}(2012){Sana}, {de Mink}, {de Koter}, {Langer},
  {Evans}, {Gieles}, {Gosset}, {Izzard}, {Le Bouquin}, \&
  {Schneider}}]{Sana-2012}
{Sana}, H., {de Mink}, S.~E., {de Koter}, A., {et~al.} 2012, Science, 337, 444,
  \dodoi{10.1126/science.1223344}

\bibitem[{{Sargent} {et~al.}(2011){Sargent}, {Srinivasan}, \&
  {Meixner}}]{Sargent-2011}
{Sargent}, B.~A., {Srinivasan}, S., \& {Meixner}, M. 2011, \apj, 728, 93,
  \dodoi{10.1088/0004-637X/728/2/93}

\bibitem[{{Schuster} {et~al.}(2006){Schuster}, {Humphreys}, \&
  {Marengo}}]{Schuster-2006}
{Schuster}, M.~T., {Humphreys}, R.~M., \& {Marengo}, M. 2006, \aj, 131, 603,
  \dodoi{10.1086/498395}

\bibitem[{{Selig} {et~al.}(2013){Selig}, {Bell}, {Junklewitz}, {Oppermann},
  {Reinecke}, {Greiner}, {Pachajoa}, \& {En{\ss}lin}}]{Selig-2013}
{Selig}, M., {Bell}, M.~R., {Junklewitz}, H., {et~al.} 2013, \aap, 554, A26,
  \dodoi{10.1051/0004-6361/201321236}

\bibitem[{{Skrutskie} {et~al.}(2006){Skrutskie}, {Cutri}, {Stiening},
  {Weinberg}, {Schneider}, {Carpenter}, {Beichman}, {Capps}, {Chester},
  {Elias}, {Huchra}, {Liebert}, {Lonsdale}, {Monet}, {Price}, {Seitzer},
  {Jarrett}, {Kirkpatrick}, {Gizis}, {Howard}, {Evans}, {Fowler}, {Fullmer},
  {Hurt}, {Light}, {Kopan}, {Marsh}, {McCallon}, {Tam}, {Van Dyk}, \&
  {Wheelock}}]{2MASS}
{Skrutskie}, M.~F., {Cutri}, R.~M., {Stiening}, R., {et~al.} 2006, \aj, 131,
  1163, \dodoi{10.1086/498708}

\bibitem[{{Smartt}(2015)}]{Smartt-2015}
{Smartt}, S.~J. 2015, \pasa, 32, e016, \dodoi{10.1017/pasa.2015.17}

\bibitem[{{Smith} {et~al.}(2023){Smith}, {Pearson}, {Sand}, {Ilyin},
  {Bostroem}, {Hosseinzadeh}, \& {Shrestha}}]{Smith_2023ixf}
{Smith}, N., {Pearson}, J., {Sand}, D.~J., {et~al.} 2023, arXiv e-prints,
  arXiv:2306.07964, \dodoi{10.48550/arXiv.2306.07964}

\bibitem[{{Soraisam} {et~al.}(2023){Soraisam}, {Matheson}, {Andrews},
  {Narayan}, {Aleo}, \& {Team}}]{Soraisam-2023}
{Soraisam}, M., {Matheson}, T., {Andrews}, J., {et~al.} 2023, Transient Name
  Server AstroNote, 139, 1

\bibitem[{{Soraisam} {et~al.}(2018){Soraisam}, {Bildsten}, {Drout}, {Bauer},
  {Gilfanov}, {Kupfer}, {Laher}, {Masci}, {Prince}, {Kulkarni}, {Matheson}, \&
  {Saha}}]{Soraisam-2018}
{Soraisam}, M.~D., {Bildsten}, L., {Drout}, M.~R., {et~al.} 2018, \apj, 859,
  73, \dodoi{10.3847/1538-4357/aabc59}

\bibitem[{{Srinivasan} {et~al.}(2011){Srinivasan}, {Sargent}, \&
  {Meixner}}]{Srinivasan-2011}
{Srinivasan}, S., {Sargent}, B.~A., \& {Meixner}, M. 2011, \aap, 532, A54,
  \dodoi{10.1051/0004-6361/201117033}

\bibitem[{{Stanway} \& {Eldridge}(2018)}]{Stanway-2018}
{Stanway}, E.~R., \& {Eldridge}, J.~J. 2018, \mnras, 479, 75,
  \dodoi{10.1093/mnras/sty1353}

\bibitem[{{Steininger} {et~al.}(2019){Steininger}, {Dixit}, {Frank}, {Greiner},
  {Hutschenreuter}, {Knollm{\"u}ller}, {Leike}, {Porqueres}, {Pumpe},
  {Reinecke}, {{\v{S}}raml}, {Varady}, \& {En{\ss}lin}}]{nifty-3}
{Steininger}, T., {Dixit}, J., {Frank}, P., {et~al.} 2019, Annalen der Physik,
  531, 1800290, \dodoi{10.1002/andp.201800290}

\bibitem[{{Stothers}(1969)}]{Stothers-1969}
{Stothers}, R. 1969, \apj, 156, 541, \dodoi{10.1086/149987}

\bibitem[{{Szalai} \& {Van Dyk}(2023)}]{Szalai2023}
{Szalai}, T., \& {Van Dyk}, S. 2023, The Astronomer's Telegram, 16042, 1

\bibitem[{{Van Dyk}(2017)}]{SVD-2017}
{Van Dyk}, S.~D. 2017, Philosophical Transactions of the Royal Society of
  London Series A, 375, 20160277, \dodoi{10.1098/rsta.2016.0277}

\bibitem[{{VanderPlas}(2018)}]{VanderPlas-2018}
{VanderPlas}, J.~T. 2018, \apjs, 236, 16, \dodoi{10.3847/1538-4365/aab766}

\bibitem[{Virtanen {et~al.}(2020)Virtanen, Gommers, Oliphant, Haberland, Reddy,
  Cournapeau, Burovski, Peterson, Weckesser, Bright, {van der Walt}, Brett,
  Wilson, Millman, Mayorov, Nelson, Jones, Kern, Larson, Carey, Polat, Feng,
  Moore, {VanderPlas}, Laxalde, Perktold, Cimrman, Henriksen, Quintero, Harris,
  Archibald, Ribeiro, Pedregosa, {van Mulbregt}, \& {SciPy 1.0
  Contributors}}]{scipy}
Virtanen, P., Gommers, R., Oliphant, T.~E., {et~al.} 2020, Nature Methods, 17,
  261, \dodoi{10.1038/s41592-019-0686-2}

\bibitem[{{Wasatonic} {et~al.}(2015){Wasatonic}, {Guinan}, \&
  {Durbin}}]{Wasatonic-2015}
{Wasatonic}, R.~P., {Guinan}, E.~F., \& {Durbin}, A.~J. 2015, \pasp, 127, 1010,
  \dodoi{10.1086/683261}

\bibitem[{{Weiler} {et~al.}(2010){Weiler}, {Panagia}, {Sramek}, {Van Dyk},
  {Stockdale}, \& {Williams}}]{Weiler2010}
{Weiler}, K.~W., {Panagia}, N., {Sramek}, R.~A., {et~al.} 2010, \memsai, 81,
  374

\bibitem[{{Wood} {et~al.}(1983){Wood}, {Bessell}, \& {Fox}}]{Wood-1983}
{Wood}, P.~R., {Bessell}, M.~S., \& {Fox}, M.~W. 1983, \apj, 272, 99,
  \dodoi{10.1086/161265}

\bibitem[{{Yamanaka} {et~al.}(2023){Yamanaka}, {Fujii}, \&
  {Nagayama}}]{Yamanaka_2023ixf}
{Yamanaka}, M., {Fujii}, M., \& {Nagayama}, T. 2023, arXiv e-prints,
  arXiv:2306.00263.
\newblock \doarXiv{2306.00263}

\bibitem[{{Yang} \& {Jiang}(2011)}]{Yang-LMC}
{Yang}, M., \& {Jiang}, B.~W. 2011, \apj, 727, 53,
  \dodoi{10.1088/0004-637X/727/1/53}

\bibitem[{{Yang} \& {Jiang}(2012)}]{Yang-SMC}
---. 2012, \apj, 754, 35, \dodoi{10.1088/0004-637X/754/1/35}

\bibitem[{{Yang} {et~al.}(2018){Yang}, {Bonanos}, {Jiang}, {Gao}, {Xue},
  {Wang}, {Lam}, {Spetsieri}, {Ren}, \& {Gavras}}]{Yang_2018}
{Yang}, M., {Bonanos}, A.~Z., {Jiang}, B.-W., {et~al.} 2018, \aap, 616, A175,
  \dodoi{10.1051/0004-6361/201832833}

\end{thebibliography}
\bibliographystyle{aasjournal}



\end{document}